\documentclass[acmlarge,screen]{acmart}

\AtBeginDocument{%
  \providecommand\BibTeX{{%
    \normalfont B\kern-0.5em{\scshape i\kern-0.25em b}\kern-0.8em\TeX}}}
    
\usepackage{multirow}

\newcommand{\quoted}[1]{{\it{``#1''}}}

\newcommand{\standpos}[0]{{{\sc Standing Position}}}
\newcommand{\collastyle}[0]{{{\sc Collaboration Style}}}
\newcommand{\layout}[0]{{{\sc Layout Locality}}}

\newcommand{\side}[0]{{{\it Side-by-side}}}
\newcommand{\face}[0]{{{\it Face-to-face}}}

\newcommand{\dcstyle}[0]{{{\it Divide \& Conquer}}}
\newcommand{\lcstyle}[0]{{{\it Loose Collaboration}}}
\newcommand{\clstyle}[0]{{{\it Close Collaboration}}}

\newcommand{\local}[0]{{{\it Local Layouts}}}
\newcommand{\distant}[0]{{{\it Distant Layouts}}}

\setcopyright{acmlicensed}
\acmJournal{PACMHCI}
\acmYear{2023} \acmVolume{7} \acmNumber{CSCW1} \acmArticle{147} \acmMonth{4} \acmPrice{15.00}\acmDOI{10.1145/3579623}

\begin{document}

\title{Side-by-Side vs Face-to-Face: Evaluating Colocated Collaboration via a Transparent Wall-sized Display}

\author{Jiangtao Gong}
\authornote{Corresponding Author}
\email{gongjiangtao2@gmail.com}
\orcid{0000-0002-4310-1894}
\affiliation{%
  \institution{Institute for AI Industry Research, Tsinghua University}
  \city{Beijing}
  \country{China}
}

\author{Jingjing Sun}
\email{jingjingxs@outlook.com}
\affiliation{%
  \institution{Institute for AI Industry Research, Tsinghua University}
  \city{Beijing}
  \country{China}
}

\author{Mengdi Chu}
\affiliation{%
  \institution{Institute for AI Industry Research, Tsinghua University}
  \city{Beijing}
  \country{China}
}
\email{mengdichu@outlook.com}

\author{Xiaoye Wang}
\affiliation{%
  \institution{Beijing University Of Technology}
  \city{Beijing}
  \country{China}
}
\email{1286357486@qq.com}

\author{Minghao Luo}
\affiliation{%
 \institution{Peking University}
  \city{Beijing}
  \country{China}
}
\email{lmh1234321@pku.edu.cn}

\author{Yi Lu}
\affiliation{%
  \institution{Beijing University Of Technology}
  \city{Beijing}
  \country{China}
}
\email{7679067@qq.com}

\author{Liuxin Zhang}
\affiliation{%
  \institution{Lenovo Research}
  \city{Beijing}
  \country{China}
}
\email{zhanglx2@lenovo.com}

\author{Yaqiang Wu}
\affiliation{%
  \institution{Lenovo Research}
  \city{Beijing}
  \country{China}
}
\email{wuyqe@lenovo.com}

\author{Qianying Wang}
\affiliation{%
\institution{Lenovo Research}
  \city{Beijing}
  \country{China}
}
\email{Wangqya@lenovo.com}

\author{Can Liu}
\affiliation{%
\institution{City University of Hong Kong}
  \city{Hong Kong}
  \country{China}
}
\email{CanLiu@cityu.edu.hk}
\renewcommand{\shortauthors}{Trovato and Tobin, et al.}
\renewcommand{\shorttitle}{Side-by-Side vs Face-to-Face}

\begin{abstract}
Traditional wall-sized displays mostly only support side-by-side co-located collaboration, while transparent displays naturally support face-to-face interaction. Many previous works assume transparent displays support collaboration. Yet it is unknown how exactly its afforded face-to-face interaction can support loose or close collaboration, especially compared to the side-by-side configuration offered by traditional large displays. In this paper, we used an established experimental task that operationalizes different collaboration coupling and layout locality, to compare pairs of participants collaborating side-by-side versus face-to-face in each collaborative situation. We compared quantitative measures and collected interview and observation data to further illustrate and explain our observed user behavior patterns. The results showed that the unique face-to-face collaboration brought by transparent display can result in more efficient task performance, different territorial behavior, and both positive and negative collaborative factors. Our findings provided empirical understanding about the collaborative experience supported by wall-sized transparent displays and shed light on its future design.
\end{abstract}

\ccsdesc[500]{Human-centered computing~Mixed / augmented reality}
\ccsdesc[500]{Human-centered computing~Collaborative and social computing devices}

\keywords{transparent display, wall-sized display, collaborative work, data manipulation}

\begin{teaserfigure}
  \includegraphics[width=\textwidth]{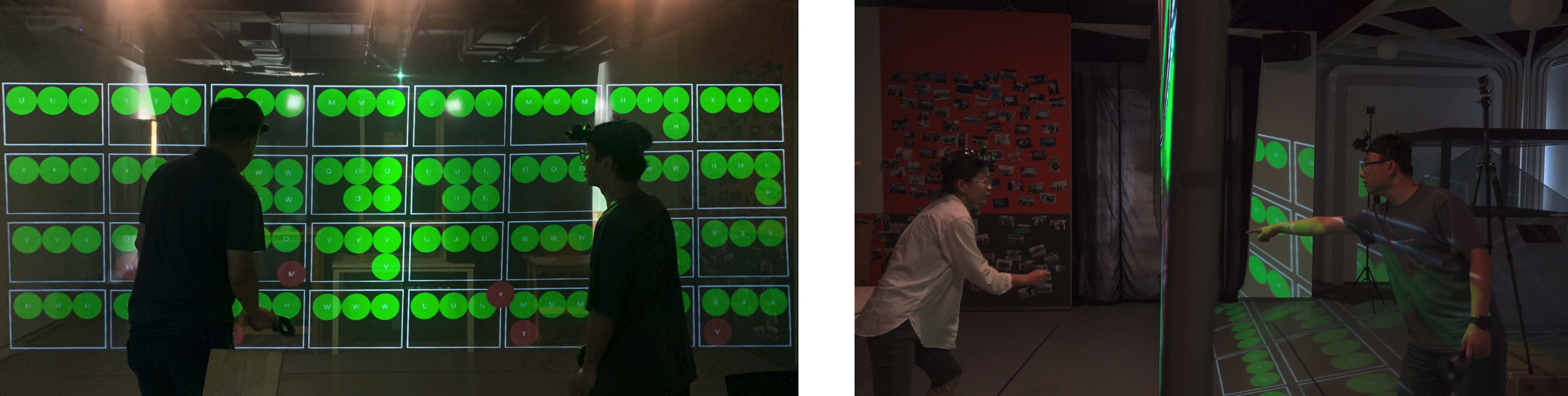}
  \caption{Two Standing Positions on a Wall-sized Transparent Display: Side-by-Side vs. Face-to-Face}
  \Description{Enjoying the baseball game from the third-base
  seats. Ichiro Suzuki preparing to bat.}
  \label{fig:teaser}
\end{teaserfigure}

\maketitle

\section{Introduction}

With the rapidly growing technology of transparent display, currently large transparent displays with high display quality can serve as an alternative platform to Head Mounted Augmented Reality Displays for providing a similar experience without introducing the social barriers between multiple users caused by headsets.

The most prominent feature of transparent displays is that it allows users to ``see-through'' it, to simultaneously view both the graphics on the screen and real-world objects and people behind the screen.
This makes transparent displays potentially highly effective in supporting face-to-face collaboration. Some researchers in HCI have investigated this. For instance, 
Gong et al.~\cite{gong2021holoboard2} designed a large-format immersive teaching board for collaborative learning. 
Transwall~\cite{heo2013transwall} explored face-to-face touch-based collaborative games design; JANUS~\cite{lee2014janus} made use of the principle of POV (Persistence Of Vision) display for readable face-to-face collaboration. 
Li et al.\cite{li2017two} used a framework of workspace awareness to investigate how a transparent display benefits colocated collaboration by enabling face-to-face collaboration and opportunistic casual interaction. 

Face-to-face collaboration via transparency metaphor medium has been shown to be effective for supporting various collaboration scenarios, including conversation~\cite{ishii1992clearboard,pejsa2016room2room}, instruction~\cite{zillner20143d} and brainstorming~\cite{kunz2010collaboard}.
Existing literature mostly studied this in a remote collaboration context by rendering the front image of a remote user on a \textit{virtual transparent display} to create a face-to-face presence. Such rendering techniques were shown to be more effective and gaze awareness for supporting collaboration compared to traditional videoconferencing. 
However, there is a lack of formal evaluation of how well colocated collaboration is supported by a large physical transparent display, particularly in comparison with a traditional large display. What are the actual differences? This paper sets out to fill this gap.



While colocated collaboration has been well-studied for interactive whiteboard~\cite{elrod1992liveboard}, traditional wall-sized display~\cite{von2016miners}, tabletop~\cite{cherek2018tangible}, it is rarely explored for transparent displays. Various methods have been used to evaluate collaborative tasks, including sensemaking~\cite{andrews2010space}, way-finding~\cite{ni2006increased}, visual analytics~\cite{reda2015effects} and searching~\cite{ruddle2015performance}. Most of these works conducted an observational study of a realistic task with an example dataset, in order to observe natural behaviors emerged in a target scenario. Such methods are great for generating real-world insights, but not as effective for comparison especially with quantitative measures because of the uncontrollable factors in working with a realistic dataset. One rare method found in the literature for quantitatively evaluating colocated collaboration was used in Liu et al.'s work~\cite{liu2014effects}. They adopted an abstract classification task used for testing the interaction efficiency of a platform at the articulatory level, making it generalizable to various data manipulation tasks. 
We found their experimental task well suited for our purpose of comparing two display setups, thus adopted it in this work.


\begin{figure}[]
\centering
  \includegraphics[width=0.7\columnwidth]{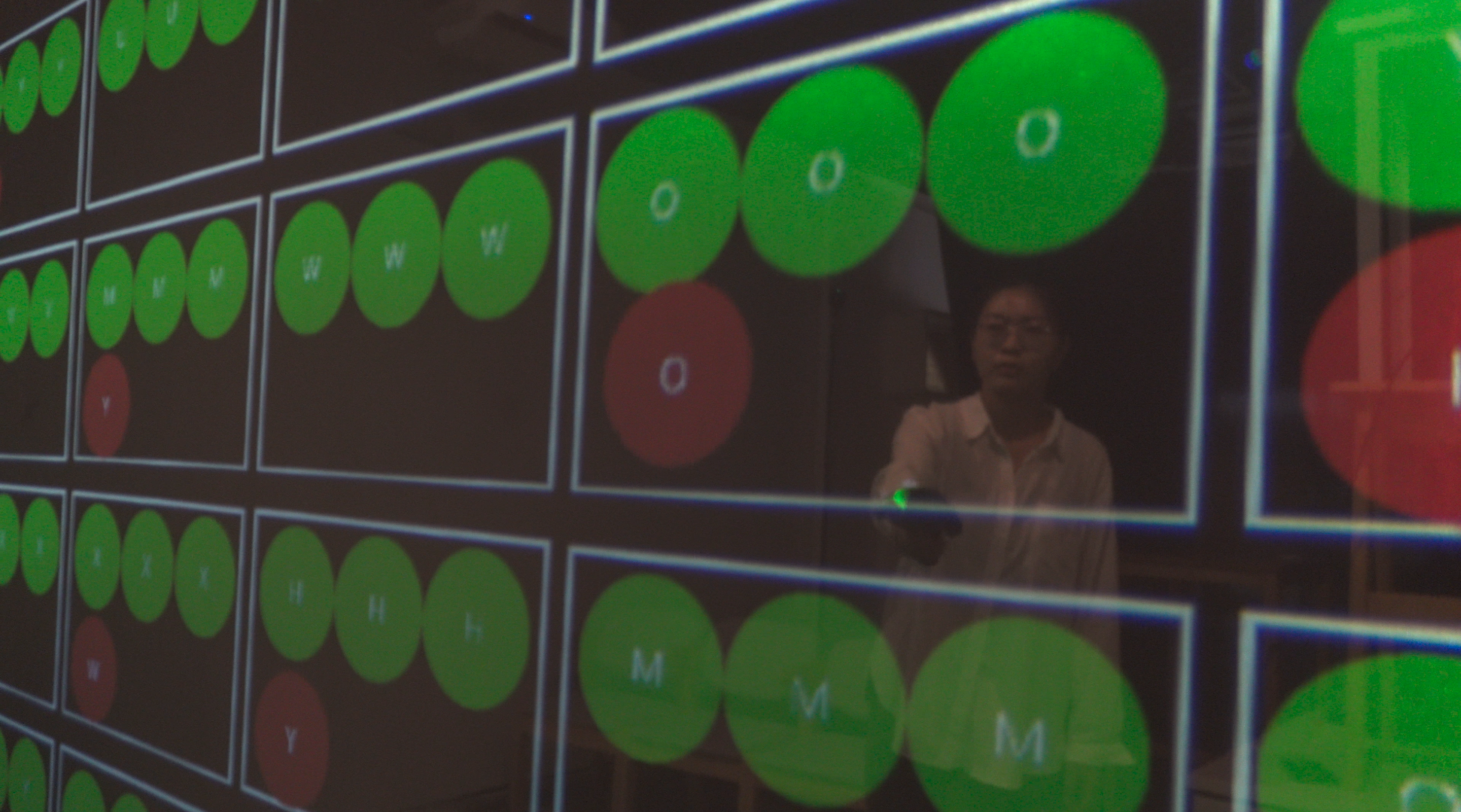}
  \caption{\textbf{View from one participant standing in Face-to-Face position.} }
  \label{fig_distance_illustration}
\end{figure}

In this paper, we present a controlled experiment to evaluate the effects of using a transparent wall-sized display for collaborative work. To investigate whether and how wall-sized transparent displays facilitate colocated collaboration, the key question is to understand the differences between the face-to-face configuration enabled by transparent displays and the side-by-side configuration offered by traditional displays. 
To control for environmental variables, we compared two conditions (side-by-side standing position and face-to-face standing position (see Fig.~\ref{fig:teaser})) using the same collaborative task on the same wall-sized transparent display.
Then, we systematically analyzed both the quantitative measures (task performance, physical navigation log data) and qualitative data (post-task interview, observation of video record). 
We found both positive factors and negative factors induced by the transparent display. Finally, we closed with a deep discussion of new territoriality and awareness issues around transparent display. 
Here we contribute to the empirical understanding of wall-sized transparent displays for collaborative purposes, adding to the repertoire of existing research on collaborative display mediums. 

Thus, the main contributions of this paper are:

\begin{enumerate}
\item First systematic empirical evaluation of a shared wall-sized transparent display with a collaborative data manipulation task.
\item Deep analysis of navigation patterns, territoriality, awareness and communication around transparent displays based on quantitative data. 
\item Further understanding about the positive and negative collaborative factors between face-to-face  and side-by-side configuration based on rich qualitative data.
\end{enumerate}


\section{Related Work}

This work builds upon existing literature on the range of empirical understanding about transparent displays, collaboration studies around shared displays, and the influence of user position in collaboration.

\subsection{Empirical Understanding of Transparent Displays}

This is a history of work related to the use of transparency metaphor in remote collaboration before physical transparent display has invented. This includes building \textit{virtual transparent displays} as whiteboards in conference that showing both drawings on the whiteboard and collaborator's shadow~\cite{tang1991videowhiteboard} or image~\cite{ishii1992clearboard}.
They conduct informal studies and observed that this kind of \textit{virtual transparent display} enables remote collaborators to share a whiteboard even more closely than if they were in the same room. The users could easily and frequently glance at each other’s faces and achieve eye contact.

Beside informal studies and observation, some comparative studies have been conducted. For instance, the 3D-Board~\cite{zillner20143d} is designed as a digital whiteboard which can display life-sized virtual embodiment's of remote collaborator. Through comparing front-facing 3D embodiment (supporting face-to-face interaction) with 2D back-facing image, they found that the results of the front-facing condition is significantly better than the back-image condition in both task performance and user experience. Room2Room~\cite{pejsa2016room2room} enabled life-size telepresence using projected augmented reality to create the experience of a face-to-face conversation. Compared with normal videoconferencing condition, Room2Room's condition got a significantly better rating on both task performance and presence. However, the efficiency and experience of remote face-to-face collaboration through \textit{virtual transparent display} still can not compare to co-located face-to-face condition.

Some of the first physical transparent screens that support colocated collaboration is based on water, smoke or fog, such as FogScreen~\cite{olwal2006immaterial} and Consigalo~\cite{olwal2008consigalo}, which are based on a wall-sized projected fog display for multi-user face-to-face collaboration. During public demonstrations of Consigalo, the authors found that the users tended to interact on the same side of screen by default, but once they saw that face-to-face is possible, they found it to to a more natural experience. This finding also has also been reported in the user study of HoloBoard~\cite{gong2021holoboard}, which is a transparent teaching board set up in a primary classroom. 

Compared to \textit{virtual transparent display}, the physical transparent display are not always transparent, and they all require a trade-off between the clarity of the graphics displayed on the screen versus the clarity of what people can see through the screen~\cite{li2017two}. 
With the development of transparent display technology, more and more clarity and high transparency displays are appearing. For example, the transparency of JANUS~\cite{lee2014janus} is around 96\% and the transparency of Tracs~\cite{lindlbauer2014tracs,lindlbauer2014collaborative} is controllable. Based on the transparency of the display, Li et al.~\cite{li2017two} and Lindlbauer et al. ~\cite{lindlbauer2016influence} conducted a comparative experiment to investigate the influence of display transparency on individual and collaborative tasks.

We can see that the previous empirical understanding of transparent display is limited, most of them have only built a prototype and conducted an informal study or observational field study. The systemic experiment studies are limited in understanding the collaborative factors of using collaborative transparent displays, especially face-to-face position and side-by-side position comparison.

\subsection{colocated Collaboration around Shared Displays}
Different from this limited understanding of transparent display, colocated collaboration has been well-studied for interactive whiteboard~\cite{elrod1992liveboard}, traditional wall-sized display~\cite{von2016miners}, tabletop~\cite{cherek2018tangible}, and spherical display~\cite{soni2021collaboration}. 

 Various methods have been used to evaluate collaborative tasks, including sensemaking~\cite{andrews2010space}, way-finding~\cite{ni2006increased}, visual analytics~\cite{reda2015effects} and searching~\cite{ruddle2015performance}. 
 Most of these works conducted observational studies using realistic tasks with example datasets to observe natural behaviors that emerged in target scenarios. For example, ethnographic approaches (such as a year-long ethnography for initial system design~\cite{wigdor2009wespace}), and numerous observational studies were conducted for understanding the collaborative work of small groups~\cite{tang1991findings}). Such methods are great for generating real-world insights but not as effective for making comparisons, especially with quantitative measures because of the uncontrollable factors in working with a realistic dataset. 
 
 Controlled experiments using quantitative methods can help systematically compare different conditions and help detect or confirm causal effects.
Cherek et al.~\cite{cherek2018tangible} conducted a study to compare reaction response times when other players triggered them by moving a tangible vs. an on-screen virtual object.
Rogers et al.~\cite{rogers2004collaborating} conducted an exploratory study to investigate the effects of the physical orientation of a display on a group working with two compared conditions: vertical versus horizontal. In the horizontal condition, group members switched more between roles, explored more ideas and had a greater awareness of what each other was doing. In the vertical condition, groups found it is more challenging to collaborate around the display. A follow-up study explored how participants, who had previous experience using both displays determined how to work together when provided with both kinds of displays.
Harris et al.~\cite{harris2009around} presented a classroom study that compared multiple-touch surfaces with single-touch for children’s collaborative learning interactions. The results of the transcript analysis of children’s interaction showed that touch condition did not affect the frequency or equity of interactions but did influence the nature of the children’s discussion.
 
From the literature, one systematic quantitative study method for evaluating colocated collaboration was used in Liu et al.'s work~\cite{liu2014effects}. Liu et al. adopted an abstract classification task used for testing the interaction efficiency of a platform at the articulately level, making it generalizable to various data manipulation tasks. 
 We found their experimental task well suited for our purpose of comparing two display setups. Therefore, we adopted it in this work. 


Motivated by the research presented in this section, this paper conducted a systematic evaluation of a transparent wall-sized display with a collaborative data manipulation task proposed by Liu et al.'s work~\cite{liu2014effects}. 


\subsection{Influence Factors of User Position in Collaboration}

The key independent variable in this work is two positions people adopt while standing in front of a transparent display ---side-by-side and face-to-face.
The user position during the collaborative activities on shared displays impacts the task performance and user experience. Chen et al. \cite{chen2020collaborative} determined the side-by-side position leads to higher task efficiency while users interact on a horizontal display. Sommer et al. \cite{sommer1969personal} found the face-to-face arrangement encourages users to have more conversations, and Whalen et al.~\cite{ha2006direct} identified the face-to-face arrangement could better support non-verbal communication. 

User position can be correlated to coupling styles when collaborating among shared displays. Tang et al.\cite{tang2006collaborative} used two observational studies to identified six coupling styles and justified how standing positions co-effect the mix-focused collaboration over the tabletop displays. Olson et al.~\cite{olson2000distance} used loose and close to justifying the compactness of coupling work, where closely coupled work refers to the high interdependent task which requires frequent communication, and loosely coupled work leads to less communication and fewer interactions. In previous coupling studies among large wall-sized displays, Jacobsen et al.~\cite{jakobsen2014up} found users are closely coupled while standing closely and talking together when looking at the same view of the large wall-sized display.
 In contrast, users are loosely coupled when standing far from each other and not talking while looking at different areas of the display. 
 
Information density (task difficulty) is another factor affecting collaborative task performance. 
Sigitov et al.~\cite{sigitov2018towards} investigated how task condition differences (focus and overview) affect collaborative activities and identified user work with small individual working areas as a focus condition, whereas working with a large working area is a so-called overview task. Prior work on collaboration typically chose focus tasks for visual analytics~\cite{grinstein2006vast,lam2011empirical,von2017giant}.



In addition, physical navigation and territoriality can be analyzed based on user position data in order to evaluate task performance. Zadow et al.~\cite{von2016youtouch} uses a camera to track user touch-related variables, such as the user’s body position and biological features like skeleton position . 
Further, Zadow et al.~\cite{von2017giant} created a movement visualization tool to indicate the different indices for physical navigation measurement. Jansen et al.~\cite{jansen2019effects} explore how moving in space and visual overview affects user perception when doing a single task under wall-sized displays. 

Prior studies on collaborative user experience also focused on user awareness, such as collaboration awareness~\cite{begole1999flexible}, situation awareness~\cite{gergle2013using}, social awareness~\cite{carroll2003notification}, context awareness~\cite{tollmar1996supporting}, workspace awareness~\cite{gutwin1996workspace}, and gaze awareness~\cite{ishii1992clearboard}. Zadow et al.~\cite{von2016miners} created a specific collaborative game to investigate users’ awareness and communication in multi-user interaction. Cherek et al.~\cite{cherek2018tangible} conducted a comparative study on tabletop display to measure each player’s awareness of these tasks by comparing reaction response times when other players triggered by moving a tangible vs. an on-screen virtual object. Lee et al.~\cite{lee2020shared} built a co-located immersive environment for collaborative data visual analytics. The user study results showed that the shared VR environment was useful in maintaining workspace awareness and to share findings with each other.


Thus, in this paper, we operationalize collaboration coupling styles and task difficulty (layout locality), and then examined task performance, how users move around the transparent display, how they establish territoriality, and how difference in standing position affect users' awareness and communication.

\section{Motivation and Research Goals}

Previous research~\cite{olwal2008consigalo,gong2021holoboard} have observed in informal user study that the users tended to interact on the same side of transparent screen(side-by-side) by default, but once they saw that face-to-face was possible, they found it to be a more natural experience. While this phenomenon have not been proved by formal experiment. Additionally, based on the existing literature and our experiences working with transparent displays, transparent displays have both properties that benefits and harms collaboration. 

On one hand, users standing on both sides of the display have a clear view of the working area on their own side, and they have more space to move around. Meanwhile, users may have a better awareness of others on the other side, as they can see each other's face, hands as well as the content being manipulated on the screen. The face-to-face configuration was shown to support engaging interaction between users in some competitive games such as boxing, tennis, etc~\cite{gong2021holoboard}. 

On the other hand, the drawbacks of transparent displays that may hinder collaboration are also apparent. Despite the name, transparent displays are not always transparent, and they all require a trade-off between the clarity of the graphics displayed on the screen versus the clarity of what people can see through the screen~\cite{li2017two}. Thus, the awareness of the opposite side may be influenced by transparency level of the display~\cite{li2017two}. The task performance may be influenced by background awareness~\cite{lindlbauer2016influence}. Furthermore, the actual display between people standing in two sides, even though a thin transparent curtain, would still interfere with collaboration by obscuring user's views of each other, distorting speech, and making it difficult to pass physical objects across the display~\cite{olwal2006immaterial,gong2021holoboard}. Therefore, it remains a question that how effective the face-to-face collaboration supported by transparent displays is, compared to the traditional side-by-side collaboration. 

This paper sets out to fill this gap by answering the following research questions:

\begin{itemize}
    \item RQ1: How does Face-to-face versus Side-by-side affect users' task performance of data manipulation?
    \item RQ2: How does Face-to-face versus Side-by-side affect users' navigation patterns and territoriality?
    \item RQ3: How does Face-to-face versus Side-by-side affect users' awareness and communication?
    \item RQ4: How does Face-to-face versus Side-by-side affect users' subjective experiences?
\end{itemize}

\section{Methodology}

The key question in investigating if and how wall-sized transparent displays support colocated cooperation is to understand the differences between the face-to-face configuration enabled by transparent displays and the side-by-side configuration offered by regular displays.
Thus, to control for environmental variables, the purpose of our study, measuring the effects of transparent display on collaboration, was implemented by comparing side-by-side and face-to-face conditions using the same collaborative task on the same wall-sized transparent display. 

Comparative study has been commonly used in social science research~\cite{collier1993comparative,ragin2014comparative}. It used both quantitative and qualitative methods to define and analyze similarities and differences caused by certain factors~\cite{lijphart1971comparative}. In our case, we used a controlled experiment with quantitative measures to enable the clearer comparison between the two main conditions. 
Meanwhile, qualitative analysis can further provide explanations for us to understand the reasons behind certain results. Given the comparative nature of our study, we needed to control the factors that are not related to display configurations yet influence task performance and user interaction patterns. Realistic tasks with realistic datasets bring many such uncontrollable factors, such as the time and actions needed for making sense of certain data points or solving a particular problem in the tasks. Therefore, we used an abstract classification task designed by Liu et al.'s prior work, which was used for comparing task performances of data manipulation on different display conditions (a wall-sized display versus a desktop computer)~\cite{liu2014effects} as well as under different collaboration strategies~\cite{liu2016shared}. 

The task was designed to eliminate as much as possible the domain-dependent intellectual work involved in a real-world task so that the differences in performance could be attributed to the main factors in the experiment. It is an established method that was also reused by other researchers in the field, such as Jakobsen and Hornbæk~\cite{jakobsen2014up} for testing the effects of locomotion on interaction with a wall-sized display.
We follow this line of work and use this task to study a new purpose of studying transparent wall-sized displays by quantitatively measuring the task performance, tracking participants' physical navigation patterns, awareness and communication over time, and analysing their post-task interviews to gain more insights. 

\subsection{Experiment Design}
The experiment adopts a $2 \times 3 \times 2$ with-in subject design with three main independent variables: \standpos{} [\side{}, \face{}], \collastyle{} [\dcstyle{}, \lcstyle{}, \clstyle{}] and \layout{} [\local{}, \distant{}]. While the task design and definition of \collastyle{} and \layout{} follows Liu et al.'s prior work~\cite{liu2016shared}, we replicate that study with a new main factor, \standpos{}, to serve our purpose. Below we briefly describe the experimental task and definitions of the factors.



\subsubsection{Task}

The task consists of moving misclassified discs into the appropriate container according to their label. Misclassified discs are colored in red, and turn green when moved into a container with a majority of discs with the same label. Participants move a disc using a pointing device that controls a cursor and a single button to pick and drop the disc with a click. The task is completed when the discs in each container have the same label, i.e. when all discs are green. 
We set up 32 containers of the same size, with up to 6 discs in each container. Each task started with a new random layout, which contained 16 misclassified red discs scattered around in containers of classified green discs. 



The original task design in Liu et al.'s work chose 8 letters to represent 8 categories of discs by labelling a letter in the center of each disc. The choices of letters were purely based on controlling even difficulties of recognizing them. We selected eight symmetrical letters ``H'', ``M", ``O'', ``U'', ``V'', ``W'', ``X'', and ``Y'' instead to ensure the same legibility on both sides of the transparent screen, while also eliminating easier-to-recognize letters such as ``A'', ``I'', and ``T''. 
We used the same font size as in their experiment - 12 pt for all the conditions to ensure the need of physical navigation (moving closer and further to the display) for completing the task.


\subsubsection{ Standing Position }
We used the same display to test two \standpos{} as illustrated in Fig.~\ref{fig_factorDesign}). Each pair of participants were required to stand in one of the two ways depending on the condition: 

\begin{enumerate}
\item \side{}: two participants stand on the same side of the transparent display to complete the task collaboratively.

\item \face{}: two participants stand on both sides of the transparent display to complete the task collaboratively.

\end{enumerate}

\subsubsection{Operationalized Collaboration Styles}

We employed Liu et al.'s operationalization of collaboration styles based on the level of coupling, to test the interaction patterns in different collaborative scenarios \cite{liu2016shared}. As illustrated in Fig.~\ref{fig_factorDesign}), three \collastyle{} were operationalized: 

\begin{figure}[]
\centering
  \includegraphics[width=0.8\columnwidth]{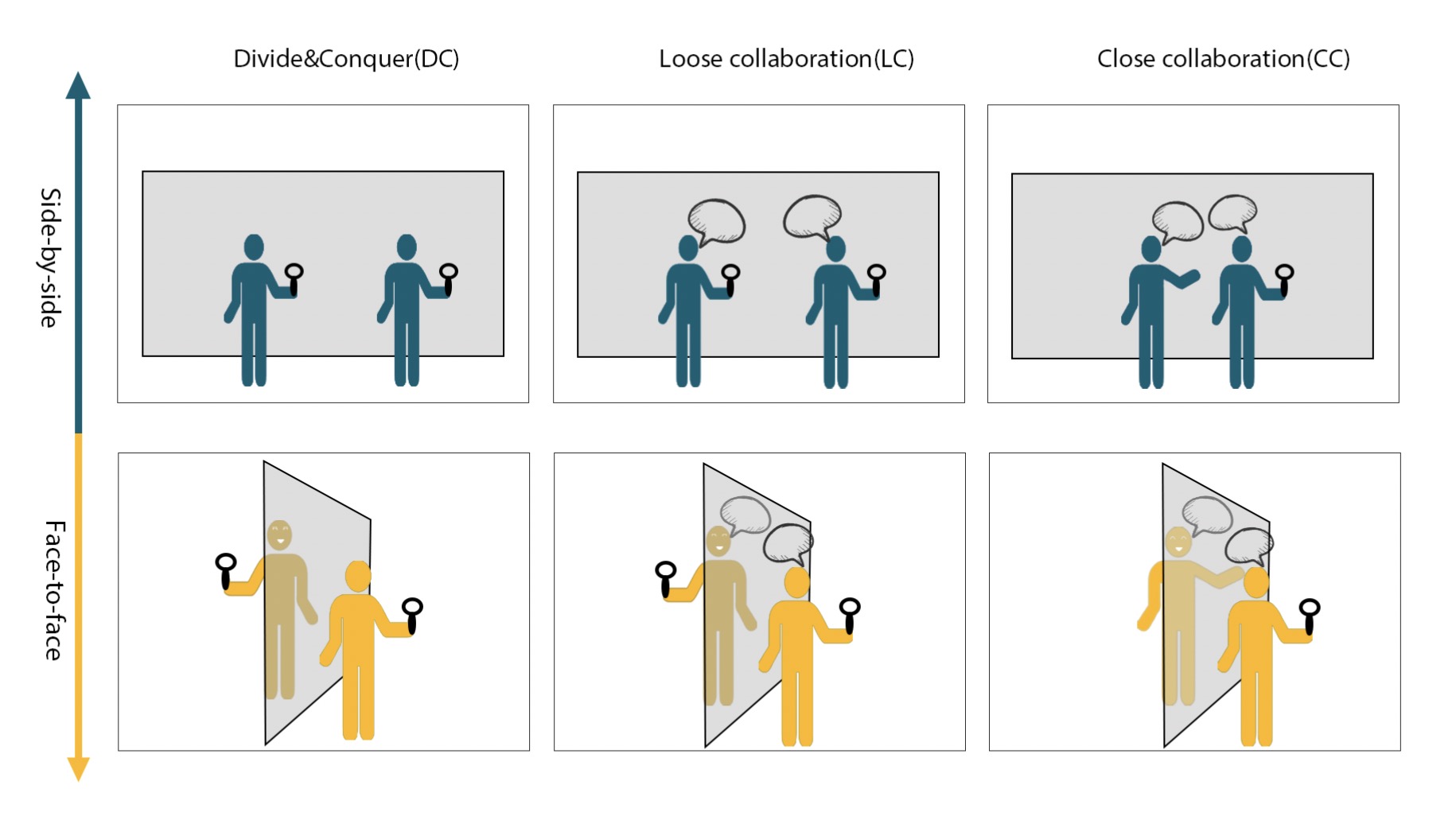}
  \caption{Operationalization of two types of standing positions (\emph{Side-by-Side} and \emph{Face-to-Face}) and three collaboration styles ( \emph{Divide\&Conquer (DC)}, \emph{Loose collaboration (LC)}, \emph{Close collaboration (CC)} }
  \label{fig_factorDesign}
\end{figure}

\begin{enumerate}
\item \dcstyle{} (DC): the task was performed by both partners in parallel. Each partner had a pointing device. Partners were not allowed to communicate with each other or help each other.

\item \lcstyle{} (LC): as above, except partners were allowed and encouraged to communicate and help each other. For example, when a partner couldn't find a container of discs with ``U'', he could say ``where is 'U' ?'', and the partner may respond ``There is a 'U'.''.

\item \clstyle{} (CC): the tasks were executed sequentially. There was only one pointing device for each pair of participants. Only one partner could perform pick and drop operations, while the other could act as an assistant to help find the target container or disc.

\end{enumerate}

\subsubsection{Layout Locality.}
We also adopted the two types of layout design in Liu et al.'s work~\cite{liu2016shared} to  operationalize the distribution of information (as well as the task difficulty) on a wall-sized display, namely \emph{Local} and \emph{Distant}:

\begin{enumerate}
\item \local{}: the target containers of all red discs were in their adjacent containers, and participants did not need to walk far to find one;

\item \distant{}: each pick-and-drop involved containers in non-adjacent columns. Participants needed to move left and right or even across the entire display to find a target container. 

\end{enumerate}

\subsection{Apparatus}
\begin{figure}[]
\centering
  \includegraphics[width=0.7\columnwidth]{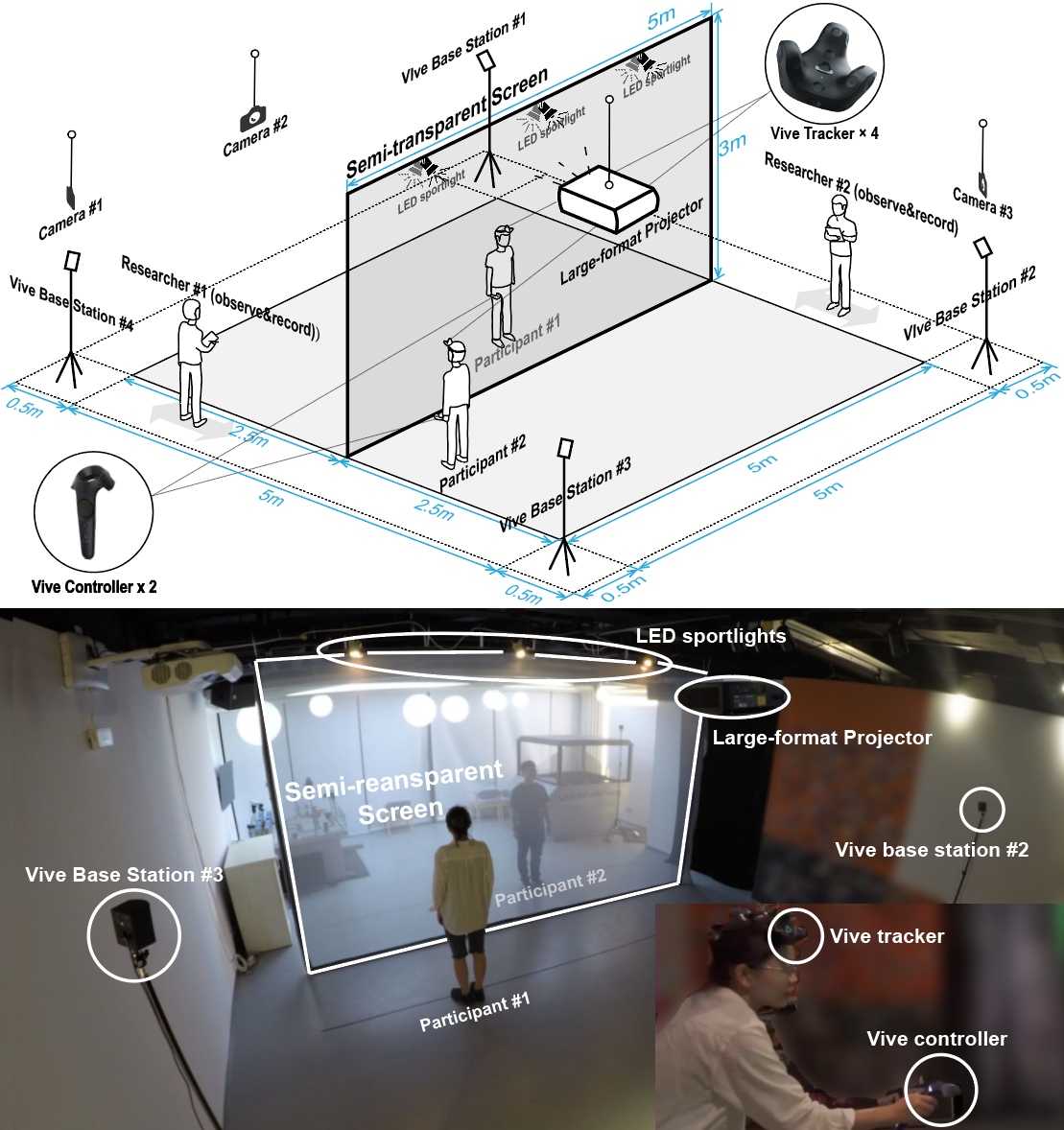}
  \caption{\textbf{Experiment Environment Setup.} }
  \label{fig_Aparatus_illustration}
\end{figure}

We used the same hardware set up as Gong et al. \cite{gong2021holoboard}. As shown in Fig.~\ref{fig_Aparatus_illustration}, we built a 5m × 3m wall-sized projected display made of translucent (semi-transparent and reflective) material to display digital material. The translucent material is constructed of glass fabric that is woven out with relatively wide holes (light transmittance is about 70\%), enabling the user to see and hear their partner when standing on the opposite side of the screen (see in Fig.~\ref{fig_distance_illustration}). Moreover, to further enhance the face-to-face collaboration, six LED spotlights were set up at the top of the display, three per side to illuminate the participant's faces so that they could see each other's facial expressions more clearly when cooperating face-to-face. 

In terms of input device and motion tracking, we used 4 SteamVR 2.0 base stations at the 4 corners of the screen, one on each corner. Two of them were in front of the projection screen while the other two were behind the screen. We asked the users to hold the HTC Vive controllers to manipulate the cursor and controller corresponding vertically. Users picked and dropped the discs via using the trigger under the handle. Two VIVE trackers were worn on their head respectively to track their head (body) positions and orientations. 

The system including the communication servers, graphic rendering, and demo applications were implemented with Unity. We divided the display area into a  4 × 8 grid with 32 rectangular containers of the same size, and each container and can holding a maximum of six discs. 


\subsection{Participants}
We used posters and flyers to recruit participants from local corporate and a local university. Thus, some of them were college students and others were corporate employees. All of our participants have normal or corrected-to-normal vision and without red/green color blindness or other mental/physical disease. In addition to the 4 participants from the pilot study to adjust study parameters, 24 participants (ages 20 - 41) were recruited for the experiment. 5 pairs were male-only, 2 pairs were female-only and 5 pairs were mixed gender. %
16 participants (8 pairs) knew each other as colleagues, others were introduced to each other before the formal experiment.	

\subsection{Procedure}
The researcher first introduced the tasks and rules to the participants and familiarised them with the conditions. A training session was conducted to get participants try out the operation and interaction with the system. 
For each type of \layout{}, participants were asked to practice for two trials. After the training session, participants were asked to start the measured trials and were instructed to complete the task as quickly as possible while minimizing errors. 

The experiment is blocked by \standpos{}, then by \collastyle{}, and final by \layout{}. In order to avoid ordering effects, the task order was fully counterbalanced across the participants. The order of \standpos{} blocks is counterbalanced through swapping, then the order of \collastyle{} is counterbalanced across pairs using a Latin Square for each \standpos{} , and the order of \layout{} is swapped for each \standpos{} and \collastyle{}.

In order to avoid accidental factors affecting the experimental results, each pair performed for each \standpos{}~$\times~$\collastyle{}~$\times$~\layout{}. For \emph{Close collaboration (CC)}, we stipulated that the pairs should exchange roles when repeating trails, to minimize effects of individual differences. In total there were 24 trials blocked by \standpos{}. 
After completing twelve trials for one \standpos{}, participants were asked to have a short break. 
After all the tasks were completed, each participant was asked to join in a ten-minutes individual semi-structured interview to talk about their subjective experiences. Two researchers observed and recorded the whole process of the experiment. Each experiment took about one hour.

\subsection{Data Collection}
In total we collected 2 (standing position) × 3 (collaboration styles) × 2 (layouts) × 2 (replications) × 12 (pairs) = 288 trails. 
Similar to previous work \cite{liu2016shared}, we implemented different data collection methods to further enrich our research findings. 
Especially noted, for two replications, we averaged the repeated data in the same condition if no accidental factor happen in two replications.

\begin{enumerate} 
\item \emph{Log Data.} 
We logged kinematic log data tracking the positions and rotations of two participant's head trackers and controllers, button clicks, and movement of discs on the screen. From the log data, we can calculated rich time indices of task performance and detailed physical navigation information, the detailed data prepossessing and definitions of measures can be seen in Appendix A.1;
\item \emph{Video Record and Observation Data.} The experiment was fully video recorded from different positions by three 4K GoPro cameras and semi-structured in-session observed by two trained researchers (see in Fig.~\ref{fig_Aparatus_illustration}). The observers noted down the number of occurrences of six types of events: (1) \emph{requested help}, one participant asking their partner for the letter being searched for; (2) \emph{effective help}, one participant finding the correct container for their partner that led to the success of the trial; (3) \emph{communication} between partners; (4) \emph{eye-contact} between partners; similar with other work~\cite{jongerius2020measurement}, we only count for reciprocal eye contact, in which two participants look at each other simultaneously. 
(5) \emph{error} as misplacing discs; (6) \emph{pick conflict}, 
both participants picking the same disc at the same time. The observation data was recorded by two researchers during the experiment independently. After each experiment, the two researchers compared their records and reached the consensus of each number of events by discussed and double-checked against the video records for confirmation. 
\item \emph{Self-report and Interview Data.} We collected 288 NASA-TLX self-report data for each participate of each condition. Besides, we collected 24 post-task interview data, which was audio recorded and fully transcribed for further coding analysis. In total, about 350 minutes of audio record were collected.
\end{enumerate} 
\subsection{Data Analysis}
\subsubsection{Quantitative analysis}
Quantitative analyses were conducted on task performance data, collaboration indices, physical navigation data and NASA-TLX scores. For the NASA-TLX scale data, we compared the side-by-side score with the face-to-face score on each item with non-parametric permutation tests. For all the other quantitative measures, we first did normally test, after that we use t-tests comparing the overall difference between \side{} and \face{} by combining six other conditions (2 \layout{} $\times$ 3 \collastyle{}) together. Then we did two-way repeated measure ANOVA for \local{} and \distant{} separately, to analyze the effects of \standpos{} and \collastyle{} as well as their interaction. After that, pairwise comparisons were performed with bonferroni corrections to compare \side{} and \face{} in each \collastyle{} (e.g. CC \& side-by-side versus CC \& face-to-face). 
Specifically, the significance level $\alpha$ = 0.05 / 3 = 0.0167 in pairwise comparison tests for three comparisons were conducted.

\subsubsection{Qualitative analysis}
We conducted post-task interviews with the participants to collect their collaborative experiences. We asked questions regarding their overall experience, their preferences regarding standing positions, collaboration styles, layout localities, and their individual and collaboration strategies during the experiment.
All the interviews were audio-recorded and transcribed. 
The interview transcripts were analyzed using the open coding method by three researchers respectively. Afterward, all codes were transcribed on sticky notes and analyzed 
with an affinity diagramming by four researchers. 


\section{Findings}

\subsection{RQ1: Task Performance}

We used ``task completion time'' (T$_{task}$) to evaluate task performance. For each trial, T$_{task}$ is the duration from picking up the first disc to dropping the last disc.
Overall we found no significant difference in T$_{task}$ between \face{} and \side{}.

\begin{figure}[]
\centering
  \includegraphics[width=1.0\columnwidth]{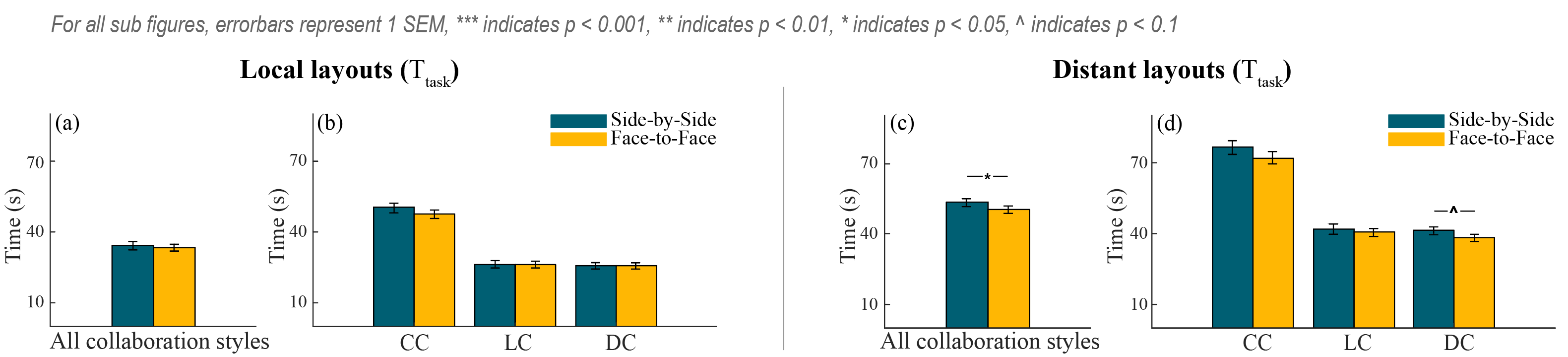}
  \caption{\textbf{Task performance.} }
  \label{fig_time}
\end{figure}

For \local{}, we found no significant difference between \side{} and \face{} on T$_{task}$ (Fig. ~\ref{fig_time}(a,b)), no matter taking three collaboration styles together or respectively. 

For \distant{}, we found it took participants less time in \face{} consistently 
on T$_{task}$ (\textit{M}$_{Face}$ = 49.9s, \textit{SD}$_{Face}$ = 5.42s, \textit{M}$_{Side}$ = 53.29s, \textit{SD}$_{Side}$ = 19.98s, \textit{F}(1,23) = 4.73, \textit{p} = 0.0401 < 0.05, \textit{$\eta^{2}$} = 0.171, Fig. ~\ref{fig_time}(c))  when combining three collaboration styles together (main effect), and no interaction effects. 

Besides, less time consumption in \face{} was reflected in DC (\textit{M}$_{Face}$ = 38.3s, \textit{SD}$_{Face}$ = 5.41s, \textit{M}$_{Side}$ = 41.32s, \textit{SD}$_{Side}$ = 6.85s, \textit{t}(23) = 1.89, \textit{p} = 0.0711 < 0.1, marginally significant, uncorrected) collaboration style on T$_{task}$~(Fig. ~\ref{fig_time}(d)).

Taking task performance results together, we found that participants consumed less time in \face{} than in \side{}, mainly with \distant{}.

\subsection{RQ2: Navigation patterns and territoriality} 
To understand user navigation patterns and territoriality, we logged participants' positions around the screen and cursor positions on screen over time. Here we introduce several indices for comparing the user interaction patterns between \side{} and \face{}, including the distance between participants, speed of navigation, and trajectory crosses between participants. 
Furthermore, we visualized participants' physical movements and cursor movements 
in the loose collaborative condition to analyze how the standing position affect territoriality while performing the same task.

\subsubsection{Distance between pairs}
We introduced two distance indices including horizontal distance and absolute distance in this section. As illustrated in Fig.~\ref{fig_distance_illustration}, we identified horizontal distance d$_{h}$ between pairs as the distance between two participants' perpendicular projection points onto the screen. Absolute distance between pairs d$_{abs}$ refers to the actual distance between pairs. We averaged the distances over time within each trial to calculate d$_{h}$ and d$_{abs}$ for each trial. 
Overall, we found pairs stood horizontally more closely in \face{} than \side{}, reflected by a smaller d$_{h}$ (\textit{M}$_{Face}$ = 1.32m, \textit{SD}$_{Face}$ = 0.68m, \textit{M}$_{Side}$ = 1.64m, \textit{SD}$_{Side}$ = 0.6m, \textit{t}(11) = 9.88, \textit{p} < 0.001).  
On the other hand, we found the absolute distance d$_{abs}$ between pairs were farther away in \face{} than \side{}, reflected by a larger d$_{abs}$ distance (\textit{M}$_{Face}$ = 2.91m, \textit{SD}$_{Face}$ = 0.36m, \textit{M}$_{Side}$ = 1.68m, \textit{SD}$_{Side}$ = 0.59m, \textit{t}(11) = -26.6, \textit{p} < 0.001).  

\begin{figure}[]
\centering
  \includegraphics[width=1\columnwidth]{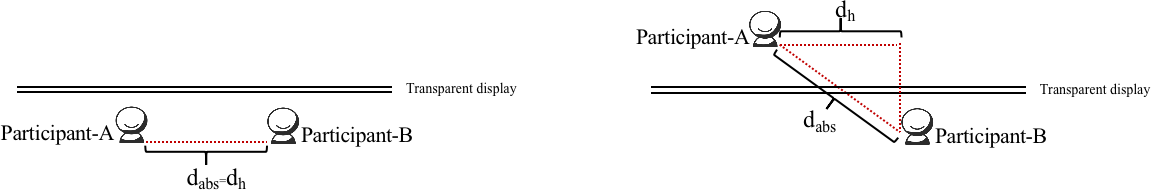}
  \caption{\textbf{Illustration of distance in side-by-side and face-to-face position.} }
  \label{fig_distance_illustration}
\end{figure}

\begin{figure}[]
\centering
  \includegraphics[width=1.0\columnwidth]{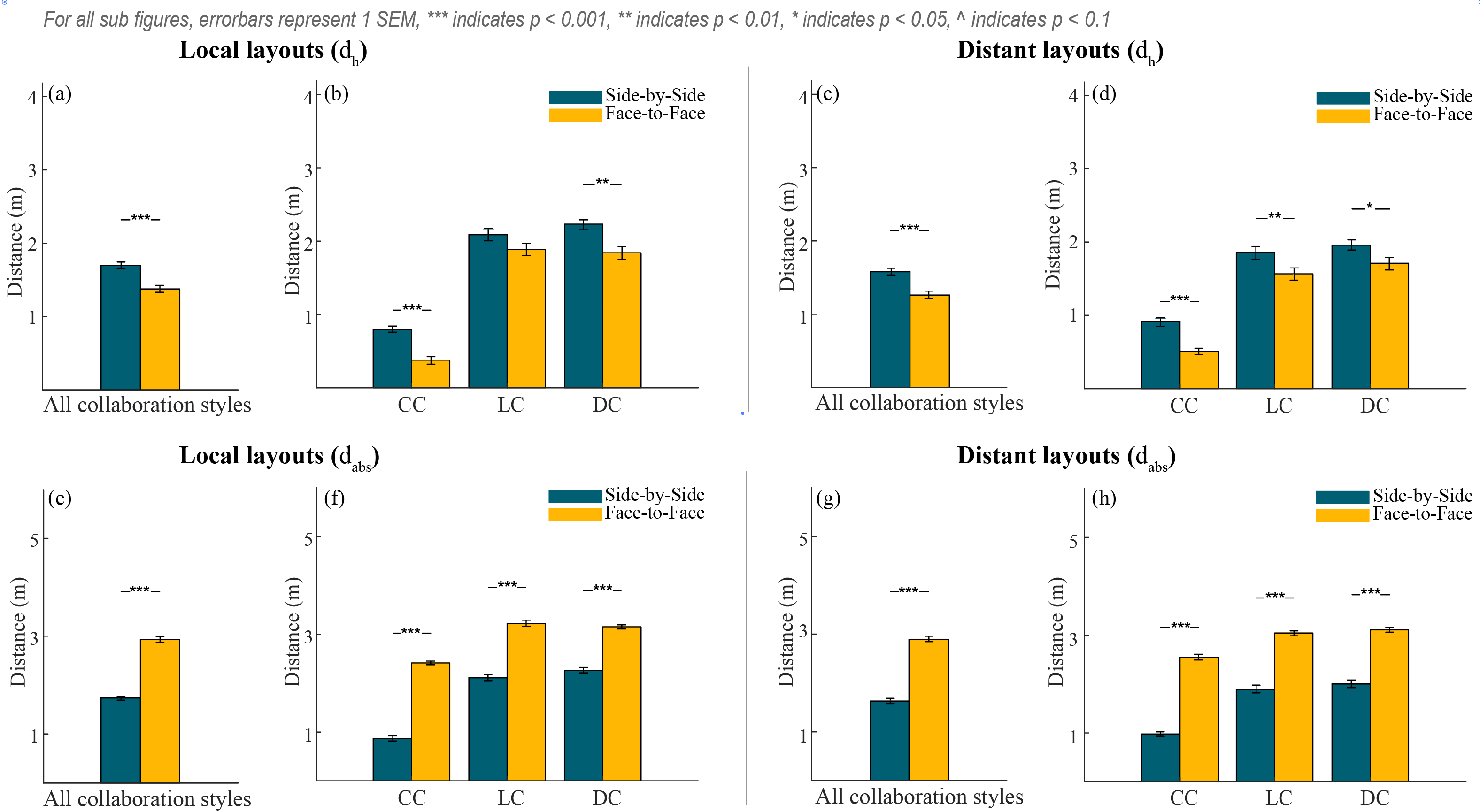}
  \caption{\textbf{Horizontal Distance d$_{h}$ (Above) and Absolute Distance d$_{abs}$ (Below) between two participates.}}
  \label{fig_distance}
\end{figure}

\textit{For local layouts}, the main effect between \side{} and \face{} is consistent with the overall T-Test. 
Participants were horizontally (d$_{h}$) closer (\textit{M}$_{Face}$ = 1.37m, \textit{SD}$_{Face}$ = 0.75m, \textit{M}$_{Side}$ = 1.7m, \textit{SD}$_{Side}$ = 0.67m, \textit{F}(1,11) = 48.4, \textit{p} < 0.001, \textit{$\eta^{2}$} = 0.815, Fig. ~\ref{fig_distance}(a)) but farther away in d$_{abs}$ (\textit{M}$_{Face}$ = 2.93m, \textit{SD}$_{Face}$ = 0.3m, \textit{M}$_{Side}$ = 1.74m, \textit{SD}$_{Side}$ = 0.66m, \textit{F}(1,11) = 407, \textit{p} < 0.001, \textit{$\eta^{2}$} = 0.974, Fig. ~\ref{fig_distance}(e)) in \face{} compared to \side{}. 
There was a significant interaction effect on d$_{abs}$ (\textit{F}(2,22) = 23.5, \textit{p} < 0.001, \textit{$\eta^{2}$} = 0.681) but no significant interaction effect on d$_{h}$.


\textit{For distant layouts}, the main effect between \side{} and \face{} is also consistent with the overall T-Test. 
Combining three collaboration styles together (main effect), we found participants were horizontally (d$_{h}$) more closely (\textit{M}$_{Face}$ = 1.26m, \textit{SD}$_{Face}$ = 0.61m, \textit{M}$_{Side}$ = 1.58m, \textit{SD}$_{Side}$ = 0.52m, \textit{F}(1,11) = 58.2, \textit{p} < 0.001, \textit{$\eta^{2}$} = 0.841, Fig. ~\ref{fig_distance}(c)) but absolutely farther away (\textit{M}$_{Face}$ = 2.9m, \textit{SD}$_{Face}$ = 0.3m, \textit{M}$_{Side}$ = 1.63m, \textit{SD}$_{Side}$ = 0.5m, \textit{F}(1,11) = 748, \textit{p} < 0.001, \textit{$\eta^{2}$} = 0.986, Fig. ~\ref{fig_distance}(g)) in face-to-face position.
There was a significant interaction effect on d$_{abs}$ (\textit{F}(2,22) = 11.5, \textit{p} < 0.001, \textit{$\eta^{2}$} = 0.510) but no significant interaction effect on d$_{h}$. 


Significant differences in pairwise comparison with in each \collastyle{} were annotated in Fig. ~\ref{fig_distance} for \local{} and \distant{} respectively. As we can observe, the main trend was that the participants were horizontally closer while absolutely farther away in face-to-face position, and this trend was consistent across different layouts and in different collaboration styles. The significant interaction effects on d$_{abs}$ show the level of differences between \side{} and \face{} were affected by \collastyle{}. From the figure we can observe that partners were working on closer or the same area of the screen (d$_{h}$) in \clstyle{} for both layouts, and \face{} allowed them to get even closer horizontally, while maintaining their physical territories separate with the transparent screen. Yet, the transparent screen did separate the partners further away in their absolute distance (d$_{abs}$) than what was necessary for comfort, and this difference was larger in \clstyle{} than in other styles for both layouts.

\subsubsection{Speed}
We introduced two speed measures, including the physical position movement speed of the participants (human speed) and the cursor movement speed (cursor speed). We calculated the human speed within each recording time interval(0.3s) by dividing the horizontal distance participant traveled within this time interval by 0.3s. Then we averaged the speed across time within one trail to get one human speed value for this trail. We calculated the cursor speed within each recording time interval (0.3s) by dividing the distance cursor trial to get one cursor speed value for this trail.

Overall, we found no significant differences on human speed or cursor speed between \face{} and \side{}.

\begin{figure}[]
\centering
  \includegraphics[width=1.0\columnwidth]{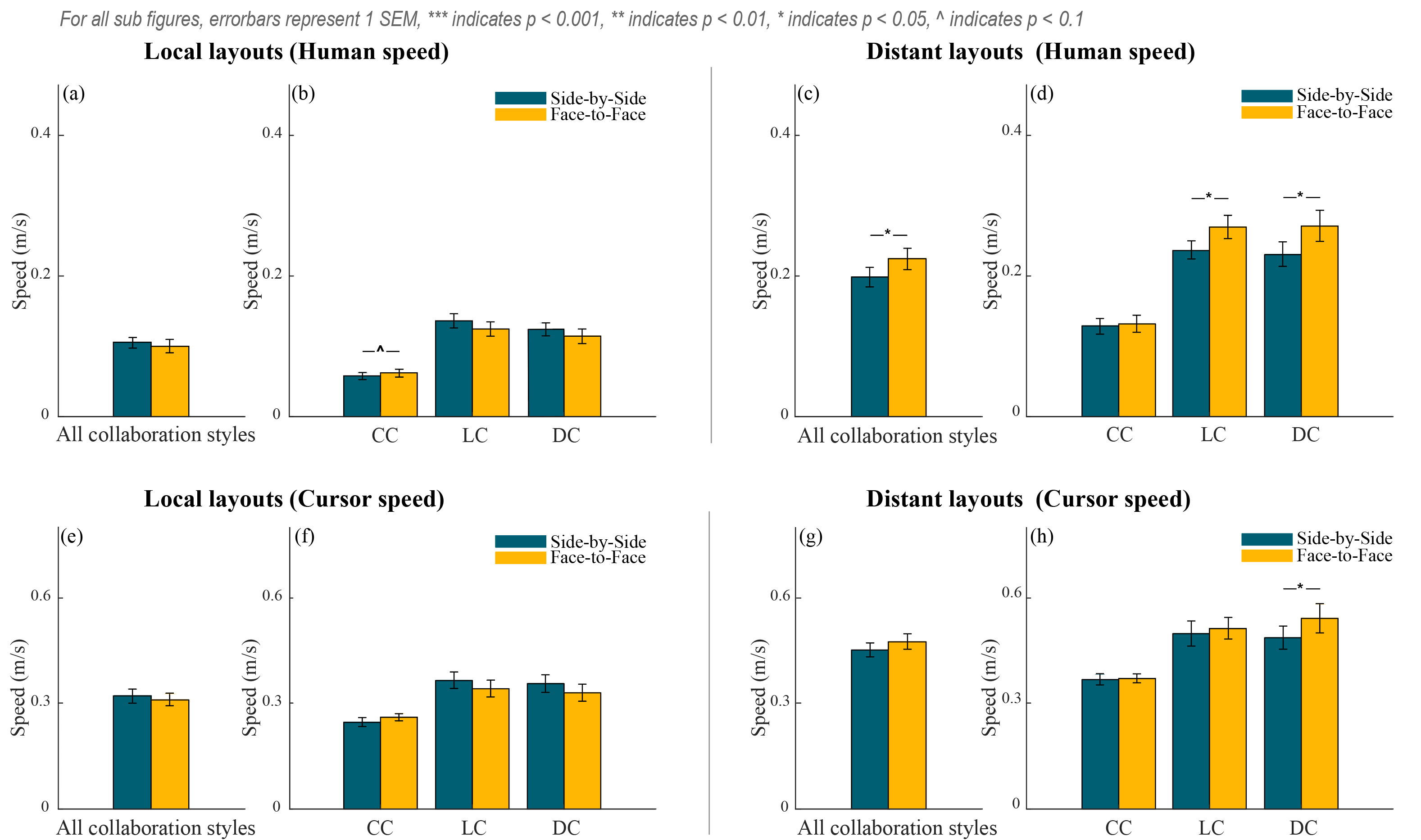}
  \caption{\textbf{Human speed (Above) and cursor speed (Below).}}
  \label{fig_speed}
\end{figure}

\textit{For local layouts}, there were significant interaction effects on human speed (\textit{F}(2,46) = 3.47, \textit{p} = 0.0396 < 0.05 , \textit{$\eta^{2}$} = 0.131) and cursor speed (\textit{F}(2,46) = 3.98, \textit{p} = 0.0256 < 0.05 , \textit{$\eta^{2}$} = 0.147). 
Specifically, we found participants moved faster in CC collaboration style (\textit{M}$_{Face}$ = 0.063m/s, \textit{SD}$_{Face}$ = 0.024m/s, \textit{M}$_{Side}$ = 0.057m/s, \textit{SD}$_{Side}$ = 0.025m/s, \textit{t}(23) = -1.72, \textit{p} = 0.0988 < 0.1, marginally significant, uncorrected), see Fig. ~\ref{fig_speed}(b).

\textit{For distant layouts}, there were significant interaction effects on human speed (\textit{F}(2,46) = 3.63, \textit{p} = 0.0343 < 0.05 , \textit{$\eta^{2}$} = 0.136) and cursor speed (\textit{F}(2,46) = 3.21, \textit{p} = 0.0495 < 0.05 , \textit{$\eta^{2}$} = 0.123). 
Combining three collaboration styles together (main effect), we found participants moved faster (\textit{M}$_{Face}$ = 0.22m/s, \textit{SD}$_{Face}$ = 0.12m/s, \textit{M}$_{Side}$ = 0.2m/s, \textit{SD}$_{Side}$ = 0.09m/s, \textit{F}(2,46) = 7.12, \textit{p} = 0.0138 < 0.05 , \textit{$\eta^{2}$} = 0.236, Fig. ~\ref{fig_speed}(c)) in face-to-face position.

Specifically, faster human movements in \face{} compared to \side{} were found in \lcstyle{} (\textit{M}$_{Face}$ = 0.27m/s, \textit{SD}$_{Face}$ = 0.1m/s, \textit{M}$_{Side}$ = 0.23m/s, \textit{SD}$_{Side}$ = 0.09m/s, \textit{t}(23) = -2.75, \textit{p} = 0.0113 < 0.0167, corrected) and \dcstyle{} (\textit{M}$_{Face}$ = 0.27m/s, \textit{SD}$_{Face}$ = 0.12m/s, \textit{M}$_{Side}$ = 0.23m/s, \textit{SD}$_{Side}$ = 0.08m/s, \textit{t}(23) = -2.47, \textit{p} = 0.0215 < 0.05, uncorrected), see Fig. ~\ref{fig_speed}(d). We also found faster cursor movement in \face{} compared to \side{} in \dcstyle{} (\textit{M}$_{Face}$ = 0.54m/s, \textit{SD}$_{Face}$ = 0.2m/s, \textit{M}$_{Side}$ = 0.49m/s, \textit{SD}$_{Side}$ = 0.18m/s, \textit{t}(23) = -2.63, \textit{p} = 0.0151 < 0.0167, corrected), see Fig. ~\ref{fig_speed}(h).

Overall, we found participants moved physically faster in \face{} than in \side{} in \distant{} when the collaboration was loose or the work was divided. Cursor movement followed the same trend but with less significant and smaller differences. 

\subsubsection{Trajectory cross}
\begin{figure}[]
\centering
  \includegraphics[width=1.0\columnwidth]{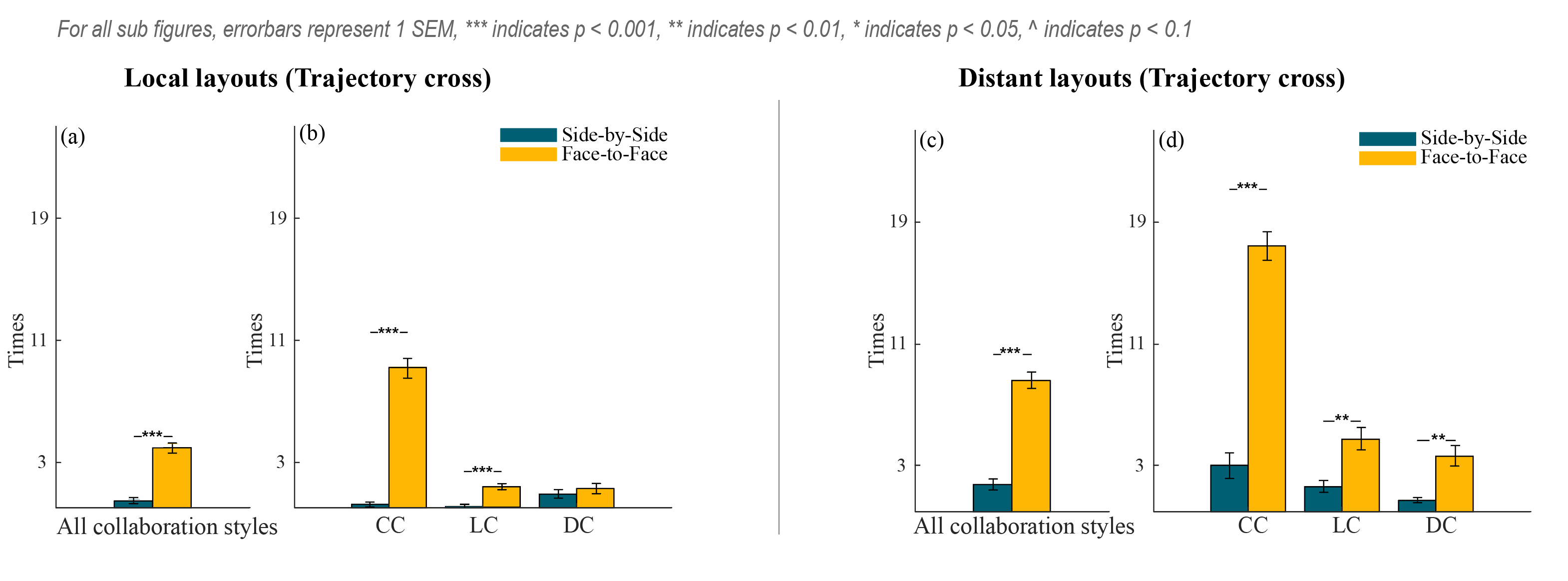}
  \caption{\textbf{Trajectory cross occurrences.}} 
  \label{fig_cross}
\end{figure}

We counted the number of occurrences in each trial that two participants' trajectories crossed horizontally. 
Overall, we found that two participants' trajectories crossed significantly more in \face{} than in \side{} (The number of times: \textit{M}$_{Face}$ = 6.27, \textit{SD}$_{Face}$ = 6.23, \textit{M}$_{Side}$ = 1.12, \textit{SD}$_{Side}$ = 2.08, \textit{t}(11) = -11.0, \textit{p} < 0.001).

Analysing \local{} and \distant{} separately, the two-way repeated measure ANOVAs yielded consistent results with the overall T-Test in both conditions. There was a significant main effect on the frequency of trajectory crossing for \local{} (The number of times: \textit{M}$_{Face}$ = 3.93, \textit{SD}$_{Face}$ = 4.15, \textit{M}$_{Side}$ = 0.43, \textit{SD}$_{Side}$ = 0.96, \textit{F}(1,11) = 113, \textit{p} < 0.001 , \textit{$\eta^{2}$} = 0.911 (see Fig. ~\ref{fig_cross}(a, c))) and \distant{} (\textit{M}$_{Face}$ = 8.61, \textit{SD}$_{Face}$ = 7.03, \textit{M}$_{Side}$ = 1.81, \textit{SD}$_{Side}$ = 2.59, \textit{F}(1,11) = 83.3, \textit{p} < 0.001 , \textit{$\eta^{2}$} = 0.883). In both layouts, participants crossed more frequently in \face{} than in \side{}.

For \local{}, pairwise comparisons within each \collastyle{} found more trajectory crosses in \face{} than in \side{} in both \clstyle{} (The number of times: \textit{M}$_{Face}$ = 9.17, \textit{SD}$_{Face}$ = 2.96, \textit{M}$_{Side}$ = 0.33, \textit{SD}$_{Side}$ = 0.31, \textit{t}(11) = -9.44, \textit{p} < 0.001, corrected) and \lcstyle{} (\textit{M}$_{Face}$ = 1.33, \textit{SD}$_{Face}$ = 0.89, \textit{M}$_{Side}$ = 0.083, \textit{SD}$_{Side}$ = 0.28, \textit{t}(11) = -5.00, \textit{p} < 0.001, corrected) (see Fig. ~\ref{fig_cross}(b)).
For \distant{}, pairwise comparisons within each \collastyle{} found more trajectory crosses in \face{} than in \side{} in every \collastyle{} (see Fig. ~\ref{fig_cross}(d)): 
\clstyle{} (The number of times: \textit{M}$_{Face}$ = 17.41, \textit{SD}$_{Face}$ = 4.1, \textit{M}$_{Side}$ = 3.04, \textit{SD}$_{Side}$ = 3.83, \textit{t}(11) = -6.84, \textit{p} < 0.001, corrected), \lcstyle{} (\textit{M}$_{Face}$ = 4.75, \textit{SD}$_{Face}$ = 3.09, \textit{M}$_{Side}$ = 1.63, \textit{SD}$_{Side}$ = 1.59, \textit{t}(11) = -3.43, \textit{p} = 0.0056 < 0.01, corrected) and \dcstyle{} (\textit{M}$_{Face}$ = 3.67, \textit{SD}$_{Face}$ = 2.2, \textit{M}$_{Side}$ = 0.75, \textit{SD}$_{Side}$ = 0.63, \textit{t}(11) = -4.13, \textit{p} = 0.0017 < 0.01, corrected). 

There were significant interaction effect between \collastyle{} and \standpos{} for both \local{} (\textit{F}(2,22) = 46.6, \textit{p} < 0.001 , \textit{$\eta^{2}$} = 0.809.) and \distant{} (\textit{F}(2,22) = 21.1, \textit{p} < 0.001 , \textit{$\eta^{2}$} = 0.657).

The trajectory crosses here indicate switches of territory in the horizontal direction, as one working at the left or right side of each other when facing the screen. Overall we can see that participants switched territories much more frequently in \face{} than in \side{}.  

\begin{figure}[]
\centering
  \includegraphics[width=1\columnwidth]{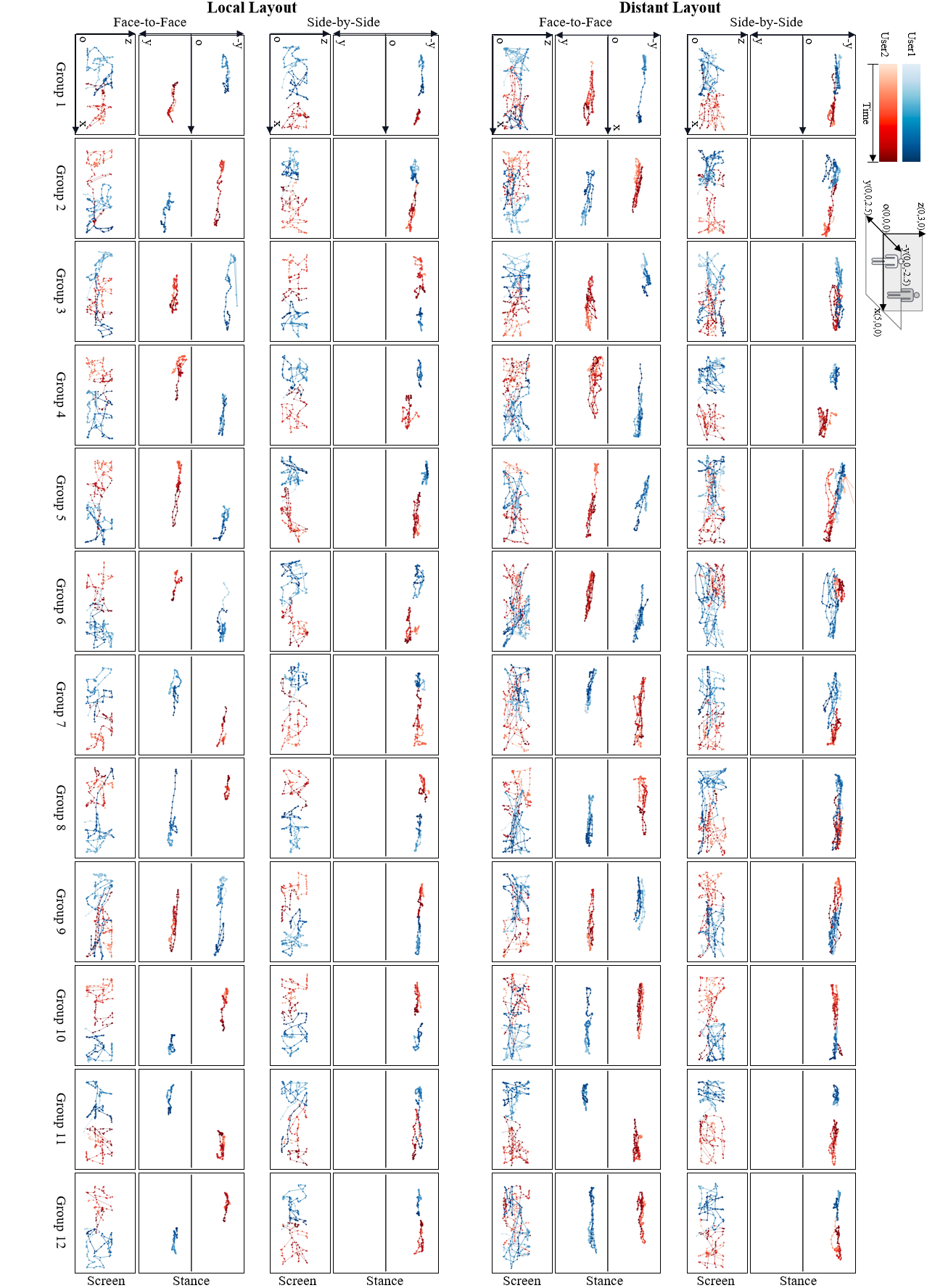}
  \caption{\textbf{Standing Position Traces and Screen Cursors Traces in Loose Collaboration} }
  \label{Traces}
\end{figure}

\subsubsection{Territoriality in loose collaboration}
While \clstyle{} and \dcstyle{} enforce or forbid participants to communicate or help each other, \lcstyle{} allows participants to work together freely, thus operationalizes a more natural situation with richer collaborative patterns. In Fig.~\ref{Traces} we visualize the accumulated trajectory patterns of both humans and their cursors for each group, to have a closer look at how \standpos{} may have affected territorialities on the screen and physically in the room respectively.  


For \local{}, 11 of the 12 groups in \side{} and 8 out of 12 groups in \face{} showed no overlap in screen work areas. This shows in both conditions, participants had strong territorial awareness and each of them worked on screen areas that were clearly delineated. We do not see large differences between \side{} and \face{} in this respect. 

For \distant{}, we can observe larger differences between \side{} and \face{} in territoriality. In \side{}, 9 of 12 groups showed a clear demarcation line dividing their work areas on the screen. Whereas in \face{}, only 2 of 12 groups had a clear separation of screen territories. The physical territories in \side{} were more mixed in \distant{} than in \local{} as the task required them to move farther to look for containers. But we can see they were less free than in \face{} for crossing to their partner's side. \side{} partners presented stronger territorial separation and a more constrained range of movements. 

This finding is consistent with the previous section on trajectory cross, showing that \face{} provided more freedom in navigating the physical space than \side{} by having partners on each side, especially in \distant{}. \face{} allowed and led to more mixture of the screen territory than \side{} while protecting their physical territories to keep it socially comfortable.

\subsection{RQ3: Users' Awareness and Communication}

In our observational data we logged the occurrences of participants' \emph{eye contact} and \emph{action conflict} (trying to pick the same disc), while other observational data didn't show significant difference between two standing position. We analyze these quantitatively to understand how \standpos{} affects partners' communication and awareness. 
Our qualitative observations also supported us to gain more insights about user awareness and how they communicated.

\begin{figure}[]
\centering
  \includegraphics[width=1.0\columnwidth]{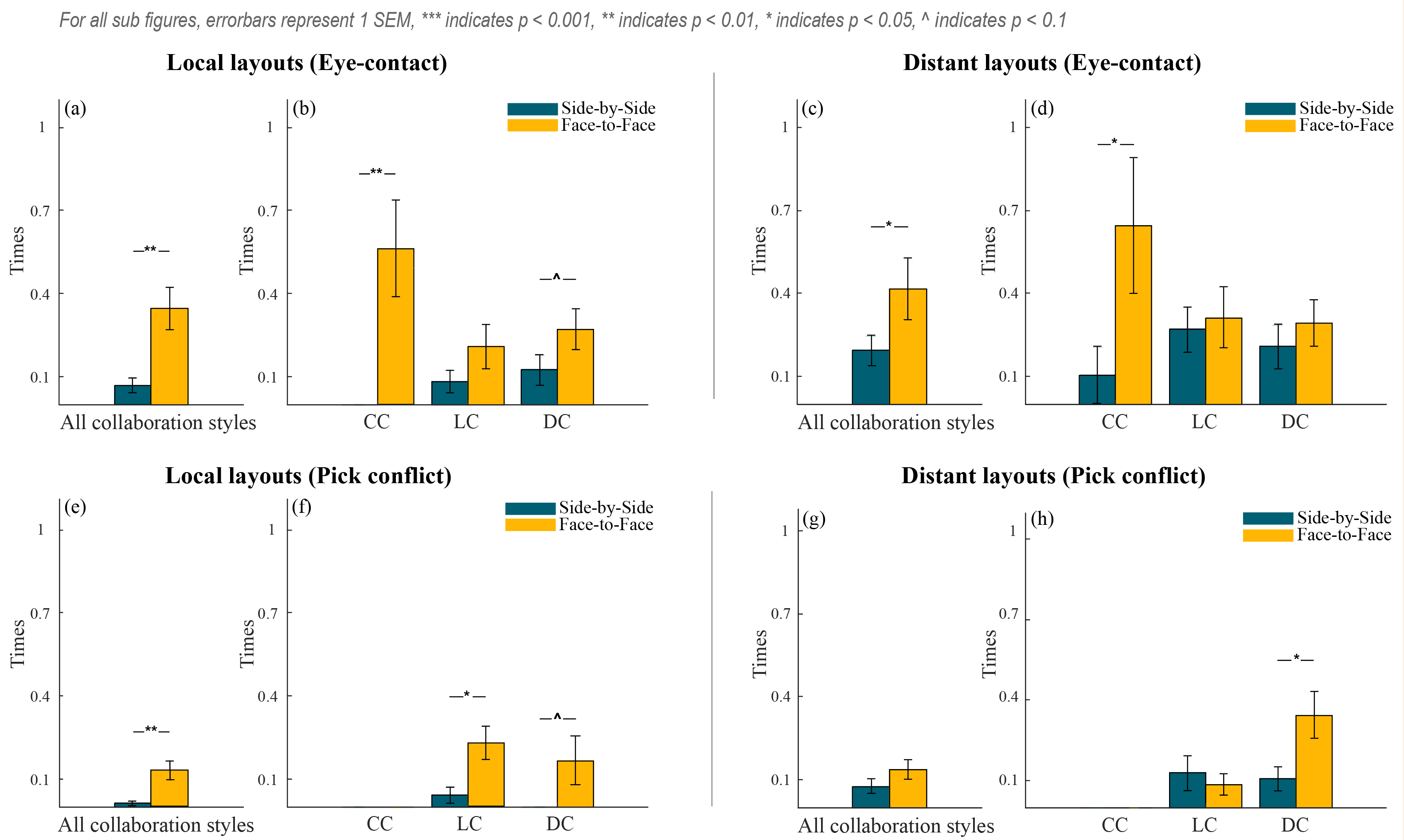}
  \caption{\textbf{Eye contact (Above) and Pick conflict (Below).}}
  \label{fig_eyecontact_pick}
\end{figure}

\subsubsection{Eye-contact}


For \local{}, combining three collaboration styles together (main effect), we found participants experienced more eye contacts (The number of times: \textit{M}$_{Face}$ = 0.35, \textit{SD}$_{Face}$ = 0.74, \textit{M}$_{Side}$ = 0.069, \textit{SD}$_{Side}$ = 0.28, \textit{F}(1,23) = 13.3, \textit{p} = 0.0013 < 0.01, \textit{$\eta^{2}$} = 0.366, Fig. ~\ref{fig_eyecontact_pick}(a)) in \face{} than in \side{}.
Within each \collastyle{}, pairwise comparisons revealed that \face{} had significantly more eye contact than \side{} in \clstyle{} (The number of times: \textit{M}$_{Face}$ = 0.56, \textit{SD}$_{Face}$ = 1.07, \textit{M}$_{Side}$ = 0, \textit{SD}$_{Side}$ = 0, \textit{t}(23) = -3.24, \textit{p} = 0.0036 < 0.01, corrected) and \dcstyle{} (\textit{M}$_{Face}$ = 0.27, \textit{SD}$_{Face}$ = 0.44, \textit{M}$_{Side}$ = 0.13, \textit{SD}$_{Side}$ = 0.39, \textit{t}(23) = -1.90, \textit{p} = 0.0695 < 0.1, marginally significant, uncorrected) collaboration styles (Fig. ~\ref{fig_eyecontact_pick}(b)).

For \distant{}, combining three collaboration styles together (main effect), we also found participants had more eye contact (The number of times: \textit{M}$_{Face}$ = 0.41, \textit{SD}$_{Face}$ = 0.92, \textit{M}$_{Side}$ = 0.19, \textit{SD}$_{Side}$ = 0.59, \textit{F}(1,23) = 6.66, \textit{p} = 0.0167 < 0.05, \textit{$\eta^{2}$} = 0.225, Fig. ~\ref{fig_eyecontact_pick}(c)) in \face{} than in \side{}.
Within each \collastyle{}, pairwise comparisons revealed that \face{} had significantly more eye contact than \side{} in \clstyle{} only (\textit{M}$_{Face}$ = 0.64, \textit{SD}$_{Face}$ = 1.33, \textit{M}$_{Side}$ = 0.1, \textit{SD}$_{Side}$ = 0.71, \textit{t}(23) = -2.17, \textit{p} = 0.0408 < 0.05, uncorrected) collaboration style (Fig. ~\ref{fig_eyecontact_pick}(d)).

There were significant interaction effects between \collastyle{} and \standpos{} on eye-contact for both \local{}(\textit{F}(2,46) = 5.02, \textit{p} = 0.0107 < 0.05 , \textit{$\eta^{2}$} = 0.179) and \distant{}(\textit{F}(2,46) = 2.88, \textit{p} = 0.0661 < 0.1, \textit{$\eta^{2}$} = 0.111). 

We can see that participants had more eye contact in \face{} condition than in \side{}, which is consist with previous research that gaze awareness could be enhanced in face-to-face position~\cite{ishii1992clearboard}. While this was true in all \collastyle{}, the difference was particularly large in \clstyle{}. Interestingly, while we see more eye contact in closer collaboration than in loose or no collaboration in \face{}, less eye contact was observed in closer collaboration in \side{} condition. We guess this might also be due to the multiple function of eye contact - both to communicate intention and to avoid or resolve conflict. In \side{}, other communication channels were used instead of eye contact, such as verbal or gestural, when participants were collaborating closely together, whereas in \face{} eye contact remained as a main communication channel. 

\subsubsection{Other communication channels}
Through our on-site observation and post-task interview, we found there are three main communication channels that participants used to communicate and cooperate:\textit{Verbal Communication}, \textit{Cursor}, \textit{Body Language}. 


\underline{Verbal Communication}: All participants reported they used verbal communication to convey messages while collaborating with their partners. In particular, P1-A shared her communication strategy: “\textit{I first told her ‘here’ and ‘there.’ But I found that was not clear enough. Later, I directed her with words like ‘below,’ and ‘above’. I think this was more efficient. For instance, if I told her 'I’m on the second line,' that helped her quickly target where I was. Two people may think differently, and the information we share can be vague. So it is more efficient to give this kind of direction.}”
 
\underline{Cursor}: Six participants indicated that they will used the cursor to direct their partners, especially when they were in \face{} position. P1-B stated that:\quoted{\textit{If it is on the opposite side, I think it was very clear when he used the cursor to indicate to me the location. But when on the same side, since the position from which he was pointing was fixed, he could not exactly point the cursor to the position I needed to go. He could only tell me 'two up' or 'two down'.}} P12-A also pointed out that she used the cursor to help her partner, even during the DC collaborative style:\quoted{\textit{although in this situation, communication is forbidden, we could still use the cursor to indicate to the other person. I think the hints were more obvious when standing on the same side.}}

\underline{Body Language}: Body language has been mentioned when participants talk about how they interacted with their partners. Three pairs of participants mentioned they used \emph{finger pointing} during collaboration. However, two teams indicated challenges of using finger pointing to guide others in \face{} condition. 
P1-B mentioned:\quoted{\textit{[In face-to-face condition], we first pointed with our fingers. But then I realized that I didn't know where she was pointing. Later I discovered that although the cursor cannot be clicked, it can still be used as sort of a guidance}}; Likewise, P10-A indicated that she barely saw her partner's finger guide while she engaged in the task:\quoted{\textit{when standing on the opposite side, I always first found the target area before I saw his finger was pointing there}.} Besides, P10-A also pointed out that she used \emph{peripheral vision} 
to check on her partner, \quoted{\textit{when I am on the same side with my partner since he is next to me, I could see him with a peripheral vision, I received this kind of physical signal more directly while in a side-by-side position.}}
   
From our observation, communication through all these channels occurred in both \side{} and \face{} conditions. However, the subjective feedback from the participants indicated different preferences for certain communication methods in different \standpos{}: cursor communication was easier in \face{} and finger pointing was harder in \face{}, while peripheral vision can be well-used in \side{}.

\subsubsection{Pick conflict}

With all the communication channels in use, let us look at the result of pick conflict, which indicates part of the effectiveness of their communication. 

For \local{}, combining three collaboration styles together (main effect), we found participants had more pick conflicts (The number of times: \textit{M}$_{Face}$ = 0.13, \textit{SD}$_{Face}$ = 0.39, \textit{M}$_{Side}$ = 0.01, \textit{SD}$_{Side}$ = 0.12, \textit{F}(1,23) = 10.27, \textit{p} = 0.0039 < 0.01, \textit{$\eta^{2}$} = 0.309, Fig. ~\ref{fig_eyecontact_pick}(e)) in \face{} than in \side{}, while no significant interaction effect was found.
Within each \collastyle{},  pairwise comparisons revealed more pick conflicts in \face{} than in \side{} happened in \lcstyle{} (The number of times: \textit{M}$_{Face}$ = 0.23, \textit{SD}$_{Face}$ = 0.47, \textit{M}$_{Side}$ = 0.04, \textit{SD}$_{Side}$ = 0.2, \textit{t}(23) = -2.58, \textit{p} = 0.0166 < 0.0167, corrected) and \dcstyle{} (\textit{t}(23) = -1.88, \textit{p} = 0.0727 < 0.1, marginally significant, uncorrected)(Fig. ~\ref{fig_eyecontact_pick}(f)). 

For \distant{}, the main effect of \standpos{} was not significant. There was a significant interaction effect on pick conflict occurrences (\textit{F}(2,46) = 5.05, \textit{p} = 0.0104 < 0.05, \textit{$\eta^{2}$} = 0.180). 
Within each \collastyle{},  pairwise comparisons revealed that more pick conflicts in \face{} than in \side{} happened in \dcstyle{} only (The number of times: \textit{M}$_{Face}$ = 0.33, \textit{SD}$_{Face}$ = 0.55, \textit{M}$_{Side}$ = 0.1, \textit{SD}$_{Side}$ = 0.31, \textit{t}(23) = -2.70, \textit{p} = 0.0129 < 0.0167, corrected) collaboration style (Fig. ~\ref{fig_eyecontact_pick}(h)). 
Notice that due to the design of this experiment, only one pick device was provided in \clstyle{} thus there was no pick conflict. 

We can see generally there were significantly more pick conflict in \face{} condition than in \side{}. Probably more pick-conflict was a side effect of more territory crossing in screen space: two participants have separate physical spaces with fuzzy personal territories while they still share a common screen space, which causes more conflict. However there was no significant difference on pick conflict between standing positions in \lcstyle{} for \distant{}. Perhaps in this condition participants were able to exercised their freedom in free collaboration and use all communication channels, which allowed them to better avoid conflicts. 

\subsection{RQ4: Users' Subjective Experience}
We analyzed the NASA-TLX scores by comparing \side{} and \face{} with non-parametric permutation tests. We found no significant difference between them in any of the scale including Mental Demand, Physical Demand, Temporal Demand, Performance, Effort and Frustration. The only thing that came close to a difference was that \face{} (M = 2.58, SD = 1.10) was more temporally demanding than \side{} (M = 2.25, SD = 1.07) with a marginal significance \textit{p} = 0.0740.

During the post-task interview, we ask users specifically about their feelings and self preference about the two collaborating standing positions. In general, 12 participants stated that they personally prefer to stand \side{} with their partners in front of the transparent display. 10 participants stated they preferred to stand on the opposite side while collaborating with their partners. There were two participants who reported their preference was based on specific circumstances. P3-B reported that she preferred \side{} collaboration in terms of collaborative experience and she preferred \face{} interaction in regard to task-performing efficiency; P10-B reported that under close and loose collaboration conditions in local layout he preferred \side{} collaboration; However, when under no collaboration condition in distance layout, he preferred \face{} collaboration. Additionally, we further coded the reasons behind their preferences and re-categorized the codes as the positive and negative factors of the two different collaboration styles., see Fig.~\ref{SVF}. We will further introduce/explain their answers in below:

\begin{figure}[]
\centering
\includegraphics[width = 1\linewidth]{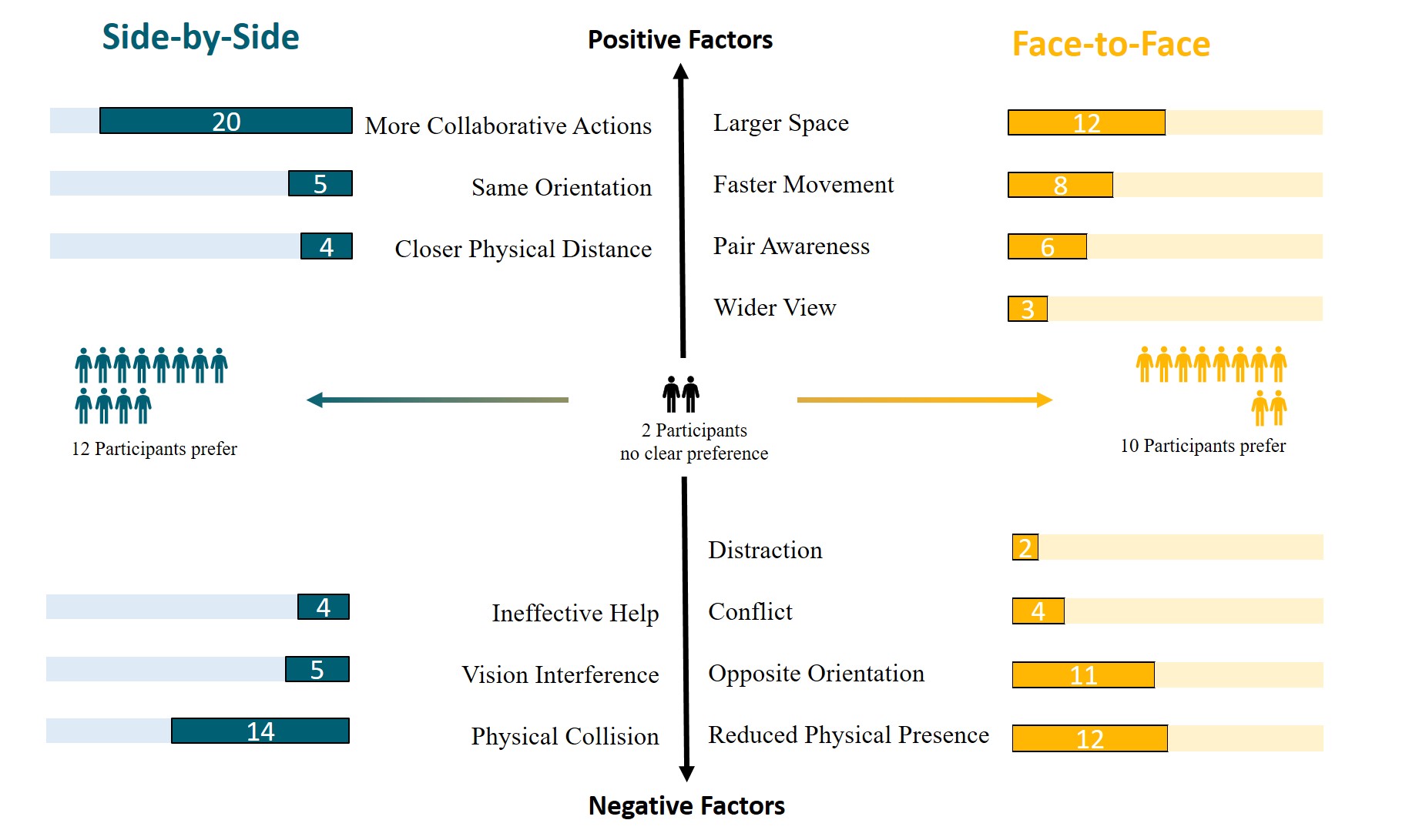}
\caption{\textbf{Collaborative Factors of Side-by-Side vs Face-to-Face}}
\label{SVF}
\end{figure}
\subsubsection{Factors affecting task performance in Side-by-Side Collaboration}

   \paragraph{Positive Factors}
   We found three main positive factors of \side{} collaboration reported by our participants as follows:
   \par
    \underline{More Collaborative Actions}:
    20 participants stated there was more collaboration involved while standing on the same side of the display. P2-B mentioned: \quoted{\textit{Standing on the same side is itself a kind of collaboration. When he couldn’t find the location, I could move to the right position and call him to come over.}}
    \par
    
    \underline{Same Orientation}:
    Five participants reported they liked the consistency of direction orientation, as P8-A explains: \quoted{\textit{...mainly because of the direction. For example, if you couldn’t find a disc, your partner could just simply say ‘it's on your left!’}}
    \par
    
    \underline{Closer Physical Distance}:
   Three participants stated that they felt physically closer to their partners. P5-A responded: \quoted{\textit{There was a sense of distance when standing on the opposite side of my partner. Since he is my partner, I felt more comfortable when he stood beside me; maybe it is a sort of a psychological effect.}}
    \par
    
\paragraph{Negative Factors}
We found three main negative factors of \side{} collaboration reported by our participants as follows:
    \par 
    
    \underline{Physical Collision}:
   14 participants reported they had concerns about colliding with their partners when standing on the same side of the transparent display (See Fig.~\ref{issue}(a)). P8-A indicated: \quoted{\textit{We happened to bump into each other at first because we hadn’t discussed a positioning strategy. I tried to avoid weaving in and out during the tasks since I didn’t want to affect my partner. However, it made it hard to get the discs that I wanted.}} P12-B also mentioned the side effects after a collision: \quoted{\textit{During the D\&C collaboration style and distant layout scenario, we inevitably had intersecting trajectories. After the running into each other, some of the disc positions we were trying to remember were suddenly forgotten. Because of the ban on communication, we could no longer recall the positions, which slowed down our completion of the tasks.}}
    \par
    
    \underline{Vision Interference}:
   Five participants reported their partner blocked their vision while switching positions (See Fig.~\ref{issue}(b)). For example, P6-B stated: \quoted{\textit{for instance, when I saw an ‘x’ on the lower right corner, but the frame was on the upper left and so I needed to move the disc and cross paths with my partner, however, because he was standing in front of the display when I passed by him, my disc would disappear and I couldn’t see it for a bit.}} 
    \par
    
    \underline{Ineffective Help}:
    Four participants pointed out that collaboration could be in effective due to different work styles. 
    P3-B mentioned that taking guidance sometimes can be troubling: \quoted{\textit{maybe we had better not have collaborated. We didn't know each other before, so I didn't know his manner or way of thinking. He didn't know mine either. For example, if there were two ‘v’s and he wanted me to place them both together in the correct frame, but maybe I preferred to drag one disc first and then on the way drag the other red disc. Maybe his way of doing something might be different from mine, so it was a little stressful when we collaborated. I wanted to choose a certain disc to move, but he might have already clicked on a different one.” P4-B added: “he thought you were going to click this ball, but you actually were going to click another. So his instruction to you is useless.}}
    \par
    
    \subsubsection{Factors affecting task performance in face-to-face Collaboration }
\paragraph{Positive Factors}
We found four main positive factors of \face{} collaboration reported by our participants as follow:

     \underline{Larger Space}:
    12 participants reported that they acquired a larger active space while standing in \face{} with their partners. P3-A mentioned the advantage of not having the bumping issue they had when standing \side{}:\quoted{\textit{when the two people stand on the same side, sometimes due to the space limitations they may try to change positions. When we were standing on different sides of the display, I didn’t worry about bumping into my partner when I changed my position.}} P1-A:\quoted{\textit{I had more freedom when standing on the opposite side. I could go further.}} This supports and explains our findings on more territorial crossing in \face{} condition.

    \underline{Faster Movement}:
    Eight participants claimed that working on opposite sides contributed to more effective individual performance since they had more space to move around and could be more focused on their own tasks. P5-B explained:\quoted{\textit{I could move faster when standing on the opposite side. I didn’t have to worry about stepping on my partner’s feet.}} This is consistent with our quantitative finding that the physical movement speed was measured higher in \face{} than in \side{} for \distant{} (see Fig. ~\ref{fig_speed}). 
    \par
    
    \underline{Collaborative Awareness}:
   Six participants mentioned that awareness of their partner rose during \face{} collaboration, which enabled even more ability to help each other in comparison to \side{} collaboration. P1-B mentions: \quoted{\textit{It was very clear for me when she pointed with the cursor.}} P4-B explained: \quoted{\textit{We had more interaction while standing on the opposite sides of the display. Maybe this is because when we communicated, he knew my relative position better. For example, when I wanted to point to him he could see me, and see whether I was in the middle or on which side. Your own spatial position was also a kind of instruction.}}
    \par
    
    \underline{Wider View}:
    Three participants reported that \face{} collaboration provided them with a broader view. P4-A mentioned that \quoted{\textit{when standing on each side, I felt my vision was wider. I could see what he was choosing, and I could help him find the next.}} (P8-A mentioned the same.) 
    
    \paragraph{Negative Factors}
We found four main negative factors of \face{} collaboration reported by our participants as follow:
     \par
     \underline{Reduced Physical Presence}:
     12 participants reported different presence issues. Five participants reported they could not see the display clearly; Five participants said they could not hear any guidance; Two participants reported they could not see the other’s body language. P9-B stated that:\quoted{\textit{I could hardly hear my partner’s instruction from the other side, and could hardly see him as well. }} P10-A mentioned \quoted{\textit{my partner didn’t talk much. When standing on the opposite side of the display, I could not discern his body language since I was also watching the content on the display. Sometimes I found he was helping point out a position, but I had already finished.}}

\par

    \underline{Opposite Orientation}:
     11 participants reported a directional issue while standing on opposite sides of the display. All complaints were the same as participant P7-A: \quoted{\textit{It was more difficult to describe when on opposite sides. It takes a lot of brain power. You need to turn your brain around to know whether it is on the 'left' or 'right', which is his 'left'.}}
    \par
    
\begin{figure}[]
\centering
\includegraphics[width = 1\linewidth]{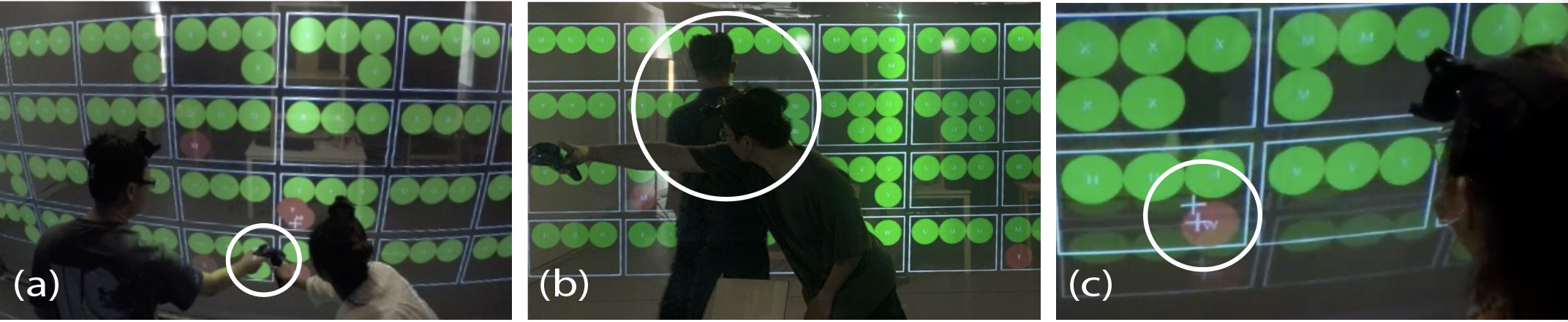}
\caption{\textbf{Examples of Collaboration Issues}: a) Body Collision (Side-by-Side); b) Vision Interference (Side-by-Side); c) Pick Conflict (Face-to-Face)}
\label{issue}
\end{figure}

    \underline{Conflict}:
    Four participants mentioned there was conflict involved. As explained by participant P3-B: \quoted{\textit{There may be minor conflict when standing on each side of the display, when there are not many discs left, both of us may go after the same disc}} (See Fig.~\ref{issue}(c)). This was consistent with our quantitative findings on Pick Conflict (see Fig. ~\ref{fig_eyecontact_pick}).
    \par
    
    \underline{Distraction}:
    Two participants mentioned that they were distracted by their partners since they needed to look both at the display and also check on their partner. P5-A stated: \quoted{\textit{when he was standing opposite me, the back light shining on him projected a shadow on the display, which affected my view.}}
    


\section{Discussion}
Previous session elaborated in detail how \side{} and \face{} affected task performance, users' navigation pattern and territoriality, communication, conflicts as well as users' subjective experiences. Here we summarize the findings and discuss them in light of existing literature. 

\subsection{Collaboration Efficiency and Experience Around Transparent Display}

From participants' subjective experience, we learned that pairs collaborated more closely in \side{} than \face{} when given the choice. However, the task efficiency was not higher but slightly lower in \side{} position. This is consistent with findings in previous work that collaboration could hinder efficiency, even though it may improve subjective experience~\cite{liu2016shared}. In our case, participants raised the issue that sometimes seeking help from their partner did not lead to effective help. Other potential reasons being indicated include more physical collision and vision interference in the \side{} configuration. On the other hand, the \face{} configuration with a see-through display clearly increased collaborative awareness, which is also known to support collaboration~\cite{ishii1992clearboard,pejsa2016room2room}. The larger space for physical movement in \face{} configuration provided a collision-free space for each participant to reach anywhere on the screen to perform tasks and help each other. This was much appreciated, but meanwhile also led to more conflicted actions as picking the same disc. 

The user preferences between \side{} and \face{} are on par. Overall there is no clear ``winner'' per se between \side{} and \face{} configuration. They are rather complimentary and have both strengths and weaknesses. Likewise, users' subjective preferences would also take consideration. 
We found collaboration has different meanings from person to person. For instance, P5-A mentioned that she “\textit{felt more comfortable}” in the \side{} position because she likes to always work closely with her partner. P1-A also mentioned that she prefers \side{} position because more collaboration has occurred. On the other hand, some participants considered task efficiency when making their preference. P3-B explained that even though collaboration is more fun, it reduces the task efficiency. P11-B indicated he prefers work by himself because the overall task was easy and not necessary for communication or collaboration. P8-B mentioned that she prefers \face{} position because she is more focused on finishing the task and she “\textit{think(s) faster when working alone}.” Interestingly, P12-B mentioned that even the goal is to work together, but she is more interested in competition. She considered the \side{} position to be more competitive, "\textit{I can obviously feel that I want to do more when standing on the opposite with my partner, I intentionally want to win the competition.}“ 

Overall, by providing two standing positions, transparent display would potentially support more diverse collaboration scenarios and user preference than traditional displays. 



\subsection{New Territoriality Around Transparent Displays}
Territoriality is widely discussed in the literature of colocated collaboration. Territories form naturally to help people mediate their social interaction through laying claim association of a space to a person during computer-supported cooperative work~\cite{ruddle2015performance, tang1991findings, scott2004territoriality, sigitov2019task}. 

The territoriality we observed in the \side{} condition was similar to findings in previous works on tabletops~\cite{scott2004territoriality} and wall displays~\cite{sigitov2019task}. 
We can see participants tended to divide the screen into separate work areas (see Fig.~\ref{Traces} (Local/Distant Layout, Side-by-Side)). This was also the case in \face{} condition with local layouts where participants did not need to search far to find a target container.
We also observed similar issues of collision and interference in \side{} situation (see Fig.~\ref{issue} (a), (b)), which is consistent with previous research indicating that people must negotiate the use of the shared space when space is scarce, because of physical interference, collisions, and occlusion problem~\cite{scott2004territoriality, ha2006direct, jakobsen2016negotiating}. 

Our findings go beyond existing knowledge about territoriality in colocated collaboration, by identifying a different territorial behaviour in \face{} configuration with a transparent large display. The large space on two sides of the screen provided the freedom for users to move around and access anywhere on the screen, which led to much more territorial overlap on the screen space compared to \side{}. In the meanwhile, the physical territories were preserved by the screen partitioning the room into two sides, to keep collaborators collision-free and socially comfortable. This finding was supported and explained by four different measures in our study comparing \face{} to \side{}, including the more overlapped cursor activity areas in the visualization of human and cursor movements (Fig.~\ref{Traces}), the significantly higher frequency of trajectory cross (Fig.~\ref{fig_cross}), the significantly faster travel speed of participants (Fig.~\ref{fig_speed}) as well as the closer horizontal distance between participants (Fig.~\ref{fig_distance}). This finding is in line with the work of Soni et al.~\cite{soni2021collaboration}, which also found that group members did not tend to partition the space when the user can move freely in front of the display.


\subsection{Awareness and Conflict Around Transparent Displays}

Awareness itself can be referred to the understanding and subsequent response to the environment and activities of others~\cite{dourish1992awareness}, which is also one of the central concepts in Computer Supported Cooperative Work.

Through quantitative analysis, we found in \face{} position,  participants commonly reported the orientation reversal problem, in which one participant’s left is the others' right. Li et al~\cite{li2014interactive}. indicated the image and text reversal issue existent with the two-sided transparent display. Our study has further understood how the mirror-effect affected people’s collaboration. 11 participants reported they experienced the confusions caused by the opposite orientation. As P7-A mentioned,"\textit{it “takes a lot of brainpower}” to converse his left to his partner’s right, with opposite orientation, muddle be one of the reasons for explaining why participants reported higher mental demand while performing tasks in the \side{} position.

We also found that as a side effect of overlapped screen territories, \side{} collaboration had more action conflicts between participants. However, this was mitigated when participants were loosely collaborating on distant layouts. Our interviews identified the communication channels participants used in all conditions, including verbal communication, body gestures and cursor indication. They also revealed insights on how cursor indication was effective and body gestures were ineffective for communication through transparent displays. People would like to use both verbal and non-verbal communication when collaboration, which consist with previous study that implicit communication is important for collaboration~\cite{che2020efficient,zheng2022coordinating,zheng2022explicit}.

In addition, consistent with previous work, we also found the lack of full transparency of the display affected the awareness and communication between partners. Some previous studies acknowledged this issue and even researched on finding feasible solutions for resolving this issue. For instance, Li et al.~\cite{li2017two} introduced a trace augmentation method to trace the user's cursor movement; Prouzeau et al.~\cite{prouzeau2018awareness} proposed Awareness Bards, Focus Map, and Step Map to assist the transition between personal and shared workspaces; Li et al.~\cite{li2016designing} and Akkil et al.~\cite{akkil2016gazetorch} tried to introduced gaze awareness for user to better collaborate for remote physical tasks. 

\subsection{Implications for Collaborative Transparent Display}
 
In this experiment, we set three collaborative styles including CC, CL,and DC and two layouts including local and distant. Critically, we found the standing position was not independent of these two variables, which means some differences between the \side{} position and the \face{} position were condition specific. This suggests some unique properties of transparent displays in certain situations and possible implications for future design. For instance, \face{} resulted in higher efficiency when users collaborated in the distant layout~(see Fig. ~\ref{fig_time}), which allowed users to have a wider view and higher physical movement speed~(see Fig. ~\ref{fig_speed}). These results suggested that the high efficiency (less time consumption) of the \face{} position could only be manifested in difficult and space-demanding tasks, while in simple tasks there was no efficiency difference between \side{} and \face{} positions. Thus, the transparent display may be utilized for maximize advantages for spatially demanding and difficult tasks where a traditional non-transparent display may fail.

We also discovered more pairs pick conflicts in the \face{} position. Throughout the post-task interview, 12 participants reported different presence issues, such as an obscured user's view of each other and distorted speech. Some presence issues can be resolved using action and speech augmentation techniques, while others cannot. For example, in the \face{}, the participates had a larger absolute distance and a smaller horizontal distance. Thus, clicking on the same object from opposite sides of the screen might be caused by a reduced physical presence. The transparent display is only a kind of "see-through" but not "walk-through" medium. Even though the transparent display is made of thin see-through fibre, it can inhibit the physical presence across the display, which causes a physical barrier.
On account of this, we suggest that the \face{} position with a transparent display could be used in more independent collaborative tasks, and the \side{} position with either non-transparent or transparent display is suitable for close collaborative tasks.

\subsection{Limitations and Future Work}

The limitations of this work are listed as follows: 

First, technical limitations of the display hardware provide limited transparency and weakened speech recognition, which may affect the performance and collaboration effectiveness of the \face{} condition. Although technical limitations may have affected some measures, we believe most of our effects and findings were strong and robust enough to hold with another similar transparent display. However, the configuration of the transparent display can be various. Different designs of transparent displays (such as different display sizes, different input devices, and different transparency levels) could be involved to further understand collaborative behaviours around transparent display in the future.

Second, our study adopted an abstract classification task to serve the purpose of making quantitative and qualitative comparisons between transparent displays and traditional one-side use displays. While providing robust and replicatable comparisons, our study sacrificed some external validity. 
Moreover, our study used a data manipulation task, thus the findings can be generalized to a range of tasks that make spatial arrangements of data items, including sense-making of scattered items, and other classification task. However, our findings may not generalize very well to other different tasks in nature. For the future work, we would conduct different collaborative tasks, such as sense-making ~\cite{morris2010wesearch} or immersive analytics~\cite{isenberg2011co} to further understand collaborative behaviours around the transparent display in other application fields.

Third, our study only implemented on pairs of people. We chose pair as the group size since it is the smallest collaboration unit size, and small group collaboration may be divided into pair collaboration. More experiments are needed in the future to understand the effect of different group sizes and to investigate the suitable group size for wall-sized transparent displays.
In the future, we will design and test other types of collaborative study to enrich our understanding about collaborative transparent display.

\section{Conclusion}
In this paper, we conducted a comparative study with an established method to investigate how wall-sized transparent displays support colocated collaborative work, in particular, how it differs from existing one-side-use wall displays. 
We achieved a rigorous quantitative and qualitative comparison by recruiting 12 pairs of participants to perform tasks face-to-face versus side-to-side using the same transparent display. 
Although previous works discussed the potential advantages and challenges of using various transparent displays to support collaboration, 
we are the first to formally evaluated the effects of transparent displays on a colocated collaboration task. 
By comparing task performances, physical navigation and territoriality, communication and conflicts, we identified unique characteristics of transparent displays: supporting new territorial behaviours with large and divided physical space; improving collaborative awareness; challenges in communication. Furthermore, we concluded a complete list of positive and negative factors identified in collaborating face-to-face and side-by-side.  
Our work contributes to the empirical understanding of transparent displays and provide insights for designing future systems on such platforms for supporting colocated collaboration. 


\begin{acks}
This project is supported by National Natural Science Foundation Youth Fund 62202267. Thanks for all the participants.
\end{acks}

\bibliographystyle{ACM-Reference-Format}
\bibliography{sample-authordraft}

\appendix
\section{Appendix of Data Processing Details}
\subsection{Data preprocessing and definitions of measures}
In our experiment, each condition was repeated twice in case of an accident. Because of no accidents happened, we average the repeated data in the same condition within each participant for all log data, recorded data and observation data.
The definition of indices derived from the log data is shown here:
\begin{enumerate}
    \item {Task completion time: }
    for each participant, the task completion time is the duration from picking up the first disc to dropping the last disc.
    \item {Horizontal distance between pairs: }
    horizontal distance between pairs refers to the distance between two participants' perpendicular projection points onto the screen. We averaged the distance across time within one trial to get one horizontal distance value for this trial. 
    \item {Absolute distance between pairs: }
    absolute distance between pairs refers to the distance between pairs in physical space. We averaged the distance of their head trackers across time within one trial to get one absolute distance value for this trial. 
    \item {Human speed: }
    we calculate the human speed within each recording time interval (0.3s) by dividing the absolute distance (participant traveled in the physical space) within this time interval by 0.3s. Then we averaged the speed across time within one trial to get one human speed value for this trial.
    \item {Cursor speed: }
    we calculate the cursor speed within each recording time interval (0.3s) by dividing the horizontal distance (cursor traveled on the display) within this time interval by 0.3s. Then we averaged the speed across time within one trial to get one cursor speed value for this trial.
\end{enumerate} 

\end{document}